\documentclass{article}
\usepackage[utf8]{inputenc}

\pdfoutput=1
\usepackage{amssymb}
\usepackage{amsmath}
\usepackage[dvips]{graphicx}
\usepackage{setspace}
\usepackage{slashed}
\usepackage{mathtools}
\usepackage{amsfonts}
\usepackage{fancyhdr}
\usepackage{xcolor}
\usepackage{graphicx}
\usepackage{rotating}
\usepackage{comment}
\usepackage{color}
\usepackage{subcaption}
\usepackage[percent]{overpic}
\usepackage{cite}
\usepackage{braket}
\definecolor{darkgreen}{rgb}{0,0.5,0}
\definecolor{darkblue}{rgb}{0,0,0.6}
\definecolor{purple}{rgb}{0.4,.2,0.7}
\newcommand{\p}{\partial}

\newcommand{\f}{\frac}

\newcommand{\be}{\begin{equation}}
\newcommand{\ee}{\end{equation}}

\usepackage[colorlinks=true,citecolor=darkgreen,linkcolor=black,urlcolor=purple]{hyperref}

\usepackage{pdfsync}

\makeatletter
\newcommand*{\defeq}{\mathrel{\rlap{%
                     \raisebox{0.3ex}{$\m@th\cdot$}}%
                     \raisebox{-0.3ex}{$\m@th\cdot$}}%
                     =} 
\makeatother

\def\be{\begin{eqnarray}}
\def\ee{\end{eqnarray}}

\newcommand{\bea}{\begin{eqnarray}}
\newcommand{\eea}{\end{eqnarray}}
\def\ben{\begin{equation}}
\def\een{\end{equation}}

\let\l=\lambda   \let\x=\xi \let\p=\phi \let\r=v

\let\f=\frac

\let\pa=\partial
\def\be{\begin{equation}}
\def\ee{\end{equation}}
\def\ba{\begin{eqnarray}}
\def\ea{\end{eqnarray}}

\newcommand{\zb}{\bar{z}}
\def\bal#1\eal{\begin{align}#1\end{align}}
\def\bs#1\es{\begin{split}#1\end{split}}

\renewcommand{\p}{\partial}

\allowdisplaybreaks  

\numberwithin{equation}{section}

\def\p{{\phi}}

\newcommand{\bz}{\bar{z}}

\def\be{\begin{equation}}
\def\ee{\end{equation}}
\def\ba{\begin{eqnarray}}
\def\ea{\end{eqnarray}}
\def\bal#1\eal{\begin{align}#1\end{align}}

\def\r{\rightarrow}

\def\f {\frac}

\def\l{\left}
\def\r{\right}

\def\x{\bar{x}}

\usepackage{tikz}
\usetikzlibrary{positioning,arrows}
\usetikzlibrary{decorations.pathmorphing}
\usetikzlibrary{decorations.markings}
\tikzset{
particle/.style={postaction={decorate}},
graviton/.style={decorate, decoration={snake, amplitude=0.8 mm, segment length=1.5 mm, pre length=0.8 mm, post length=0.8 mm}},
photon/.style={
        decoration={complete sines, amplitude=0.15cm, segment length=0.2cm},
        decorate    
    },
gluon/.style={
        decoration={coil, aspect=0.75, mirror, segment length=1.5mm},
        decorate
    }
}
 
\def \be {\begin{equation}}
\def \ee {\end{equation}}

\newcommand{\dsq}{\vec{\partial}^{\, 2}}

\renewcommand{\p}{\partial}

\newcommand{\blambda}{\bar{\lambda}}

\usepackage{framed}

\begin{document}
\onehalfspacing

\begin{center}

~
\vskip5mm

{\LARGE  {
Focusing bounds for CFT correlators and the $S$-matrix \\
\ \\
}}

\vskip10mm

Thomas Hartman,$^1$\  Yikun Jiang,$^{1}$\ Francesco Sgarlata,$^{1}$\  and Amirhossein Tajdini$^{2}$

\vskip5mm

{\it $^1$ Department of Physics, Cornell University, Ithaca, New York, USA
} 
\vskip5mm
{\it $^2$ Department of Physics, University of California, Santa Barbara, CA, USA } \\

\vskip5mm

\end{center}

\vspace{4mm}

\begin{abstract}
\noindent
The focusing theorem in General Relativity underlies causality, singularity theorems, entropy inequalities, and more. In AdS/CFT, we show that focusing in the bulk leads to a bound on CFT $n$-point functions that is generally stronger than causality. Causality is related to the averaged null energy condition (ANEC) on the boundary, while focusing is related to the ANEC in the bulk. The bound is derived by translating the Einstein equations into a relation between bulk and boundary light-ray operators. We also discuss the consequences of focusing for the flat space $S$-matrix, which satisfies a similar inequality, and give a new derivation of bounds on higher derivative operators in effective field theories.  The string theory $S$-matrix and CFT correlators in conformal Regge theory also satisfy the focusing bound, even though in these cases it cannot be derived from the standard focusing theorem.

 \end{abstract}

\pagebreak
\pagestyle{plain}

\setcounter{tocdepth}{2}
{}
\vfill
\tableofcontents

\newpage

\date{}

\section{Introduction}

Causality imposes powerful constraints on effective field theories and serves as a source of intuition for general properties of asymptotic observables. In asymptotically flat spacetime, causality is tied to analyticity and unitarity of the $S$-matrix. This is a subject with a long history (e.g., \cite{Eden:1966dnq,Adams:2006sv}), but a recent hightlight was the discovery that causality  completely fixes the tree-level coupling of three gravitons, picking out Einstein gravity as the unique consistent possibility at low energies, and points toward a string-like tower of higher spin particles as the only path to a UV completion \cite{Camanho:2014apa}. Causality in asymptotically flat spacetime also helped to motivate the chaos bound \cite{Maldacena:2015waa} as well as new dispersive sum rules for the $S$-matrix that lead to rigorous two-sided bounds on Wilson coefficients  \cite{Arkani-Hamed:2018ign,Arkani-Hamed:2020blm,Caron-Huot:2020cmc,Bellazzini:2020cot,  Tolley:2020gtv,Bern:2021ppb,Caron-Huot:2021rmr,Bern:2022yes,Caron-Huot:2022jli}.

In asymptotically anti-de Sitter spacetime, causality is tied to analyticity and unitarity of correlation functions in the dual CFT. In the limit of large impact parameter, causality in the bulk implies the averaged null energy condition (ANEC) on the boundary \cite{Hofman:2008ar,Kelly:2014mra}. This has been understood directly from the dual CFT \cite{Hartman:2015lfa,Hartman:2016dxc,Hofman:2016awc,Faulkner:2016mzt,2017JHEP...07..066H}, and is now known to be part of a larger set of constraints derived from dispersive sum rules for CFT correlators  (e.g, \cite{Afkhami-Jeddi:2016ntf,Belin:2019mnx,Caron-Huot:2020adz,Caron-Huot:2021enk}).  

In semiclassical general relativity, causality is often viewed as a consequence of a more basic inequality: the averaged null energy condition satisfied by matter fields, $\int du T_{uu} \geq 0$ \cite{Gao:2000ga,Graham:2007va}. The ANEC is an input to causality theorems proved by Tipler \cite{Tipler:1976bi,Tipler:1978zz}, Hawking \cite{Hawking:1991nk}, and Gao and Wald \cite{Gao:2000ga}, among others.  To be clear, this is different from the CFT ANEC which was studied in the literature just cited.  The CFT ANEC is dual to bulk causality, but the bulk ANEC is a different inequality, related to a different positive operator in the dual CFT.

The ANEC satisfied by matter fields in a gravitational theory is also the key assumption to prove a number of other fundamental results, including the focusing theorem, the second law of black hole thermodynamics, and topological censorship \cite{Graham:2007va}. From this point of view, it seems that perhaps the bulk ANEC is a fundamental ingredient to understand positivity in the $S$-matrix and CFT.

Our goal in this paper is to explore how the bulk ANEC, focusing, and causality are encoded and interrelated in asymptotic observables.  To illustrate the main ideas, let us consider the classical scattering of a high energy probe through a gravitational shockwave. In the null coordinates $(u=t-y$, $v=t+y$, $\vec{x})$, a shockwave geometry is produced by a highly boosted particle moving in the $v$-direction, with a stress tensor localized at $u=0, \vec{x} = 0$. Probe particles that cross the shockwave experience a time delay $\Delta v(\vec{x})$, with $\vec{x}$ the impact parameter. Causality requires the time delay to be non-negative:
\begin{align}
\Delta v(\vec{x}) \geq 0 \ .
\end{align}
In asymptotically flat spacetime, for which one can define scattering amplitudes, the time delay is proportional to the derivative of the phase shift,  $\p_s\delta(s,\vec{x})$, so causality requires this to be non-negative in the semiclassical regime.  

In general relativity, with or without higher curvature corrections, the equations of motion imply
\begin{align}\label{dvTintro}
\dsq \Delta v = - 16 \pi G_N \int_{-\infty}^{\infty} du (T_{uu}+T^{\rm grav}_{uu})  \ ,
\end{align}
where $T_{\mu\nu}$ is the matter stress tensor and $T_{\mu\nu}^{\rm grav}$ is the pseudotensor that accounts for graviton energy. The Green's function for the operator $\dsq$ is negative, so it follows immediately from \eqref{dvTintro} that the ANEC, in the form $\int du(T_{uu}+T_{uu}^{\rm grav}) \geq 0$, implies causality. This is an instance of the Gao-Wald theorem. Under certain conditions --- but not in general \cite{Engelhardt:2016aoo} --- the relation can also be inverted to show that causality implies the ANEC. 

Using this, we will show that the ANEC implies constraints on higher derivative operators in EFT, including the uniqueness of the Einstein gravity 3-point vertex and the other bounds derived in \cite{Camanho:2014apa}. Higher curvature coupling constants enter through the graviton pseudotensor, $T_{\mu\nu}^{\rm grav}$.

\begin{figure}
\begin{center}
\begin{overpic}[width=4in,grid=false]{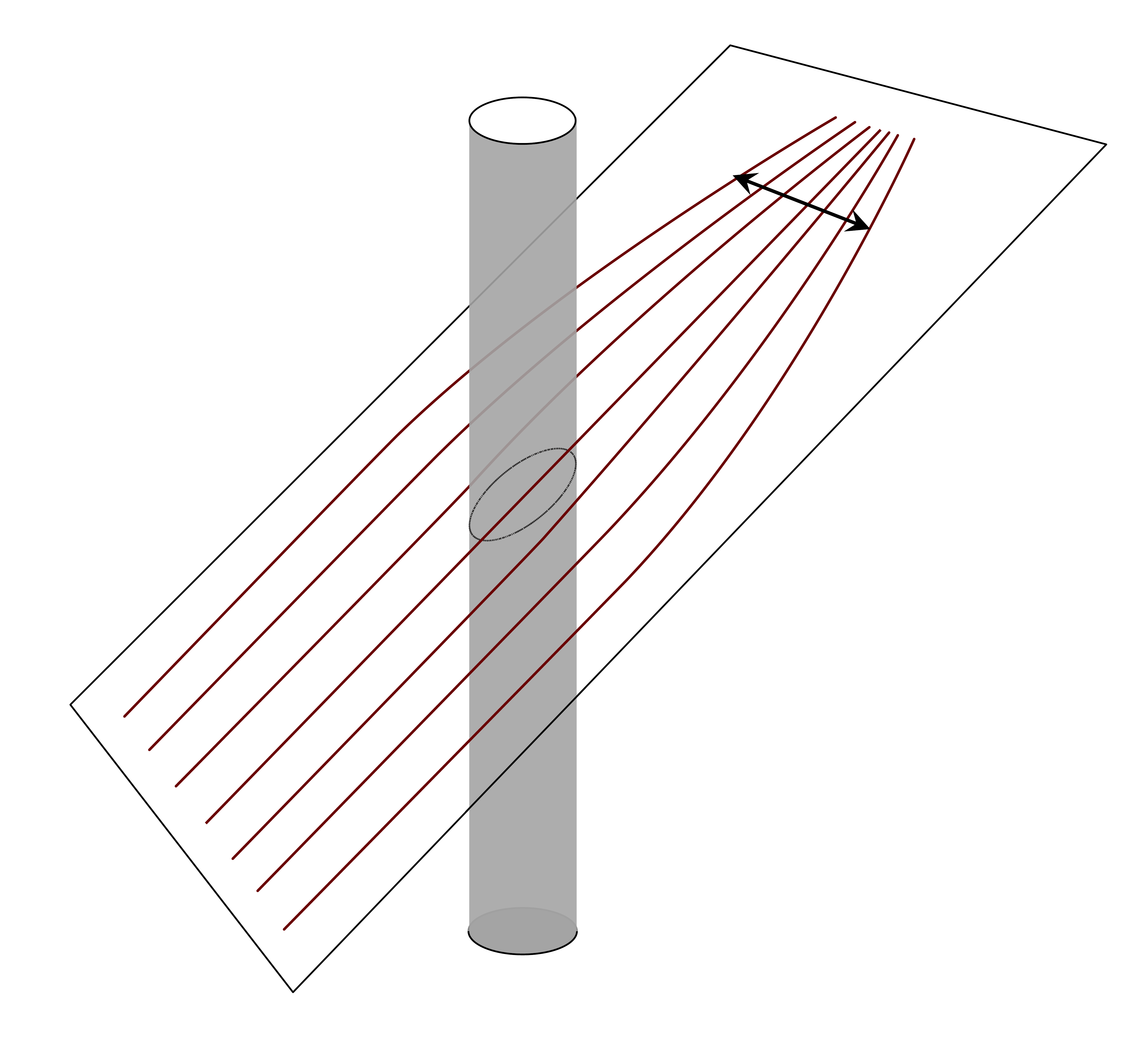}
\put (52,10) {\parbox{.6in}{target  worldline}}
\put (77,68) {Area}
\end{overpic}
\end{center}
\caption{Transverse focusing of high-energy particles scattering through a target. Focusing implies that the $(D-2)$-dimensional transverse area element $\delta A$ is non-increasing at late times. (The particles also have a time delay, but the figure is drawn in a gauge where the time delay is not visible.)\label{fig:focusing}}
\end{figure}

By \eqref{dvTintro}, the ANEC implies the differential bound
\begin{align}\label{focusIntro}
\dsq \Delta v \leq 0 \ .
\end{align}
We refer to this as the `focusing bound'. As we will explain in section \ref{ss:focusing}, it is equivalent to the integrated focusing theorem in this setting. In fact, it is a special case of Hawking's area theorem, applied to a causal horizon in the shockwave geometry. The left-hand side is proportional to $\frac{d}{du} \delta A$, where $\delta A$ is the cross-sectional area of a family of high-energy probes that begin on parallel null rays, after passing through the shock. Therefore \eqref{focusIntro} is the statement that high-energy scattering causes particles to focus rather than defocus. See Figure \ref{fig:focusing}.

The $S$-matrix version of the focusing bound is
\begin{align}\label{focusIntroDelta}
\p_s \dsq \mbox{Re}\, \delta(s,\vec{x}) \leq 0  
\end{align}
at large $s$.
The derivation of this bound from the ANEC or the focusing theorem only applies to spin-2 theories (general relativity plus higher curvature corrections). It is an open question whether it holds in general. We are not aware of any counterexamples in consistent theories.  It is violated in a theory of an isolated massive higher spin particle, but such theories are known to be inconsistent for other reasons\cite{Velo:1969txo,Porrati:1993in,Cucchieri:1994tx,Camanho:2014apa, Arkani-Hamed:2017jhn, Afkhami-Jeddi:2018apj,Bellazzini:2019bzh}. It is satisfied by the Virasoro-Shapiro amplitude in the appropriate regime, so it is tempting to speculate that the focusing bound \eqref{focusIntroDelta} is a required property of the tree-level $S$-matrix in any consistent theory of gravity. 

We also derive a parallel set of results for asymptotically anti-de Sitter spacetime. The bulk ANEC is a CFT operator that is manifestly positive, so the CFT derivation of the bulk ANEC is much simpler than the CFT derivation of the boundary ANEC in \cite{2017JHEP...07..066H}. The analogue of \eqref{dvTintro} in AdS is a differential equation relating bulk light-ray operators to boundary light-ray operators.  This turns out to be a very convenient way to encode the bulk Einstein equation in the dual CFT. We show how the bulk ANEC operator controls the Regge OPE of probe operators in the CFT, and use this to derive constraints on CFT 4-point functions in the Regge limit. Most of these results are limited to graviton exchange in the bulk, but we also test the CFT focusing inequality in conformal Regge theory and find that it continues to hold, through a somewhat surprising conspiracy of various contributions from saddles and poles. 

We will only consider small excitations (and shockwaves) above the AdS vacuum. Other circumstances in which the bulk null energy condition plays a fundamental role include black holes, where shockwaves are related to chaos and the butterfly effect \cite{Shenker:2013pqa, Shenker:2013yza, Roberts:2014isa}, and holographic RG flows, where the NEC is directly responsible for the $C$-theorem in any number of dimensions \cite{Myers:2010xs, Myers:2010tj}. See also \cite{Caron-Huot:2022lff} for recent work on thermal shockwaves which uses related methods to study the Einstein equations in the dual CFT. Perhaps the bulk ANEC can also be understood more generally in terms of operator growth, as in 2D gravity, where the bulk ANEC operator is dual to the `size' operator in the SYK model \cite{Susskind:2018tei, Brown:2018kvn, Lin:2019qwu}. Ultimately, it would be of great interest to prove (or disprove) the quantum focusing conjecture \cite{Bousso:2015mna} from this point of view, since in many cases entropy bounds and other features of the bulk reconstruction map can be traced back to quantum focusing \cite{Akers:2016ugt}.

\bigskip

\noindent The outline and a slightly more technical summary of the main results are as follows:

\begin{itemize}
\item We discuss the relation between the $S$-matrix focusing bound \eqref{focusIntro} and the standard focusing theorem (section \ref{ss:focusing}).
\item We show that the CEMZ causality constraints on 3-point vertices \cite{Camanho:2014apa} can all be derived from the bulk ANEC (sections \ref{ss:cemz}-\ref{ss:cemzexample}).
\item We show that the Virasoro-Shapiro amplitude obeys the focusing bound \eqref{focusIntroDelta}, despite the fact that the usual focusing theorem does not apply (sections \ref{ss:relateS}-\ref{ss:higherspin}).
\item We derive the bulk ANEC for free scalar fields in AdS, and write the bulk ANEC operator $\int du T_{uu}^{\rm bulk}$ as a manifestly positive operator in the dual CFT (section \ref{s:anechkll}).
\item We translate the Einstein equation (or focusing equation) into a CFT operator equation, which relates a bulk light-ray operator $\int du T_{uu}^{\rm bulk}$ to the boundary light-ray operator $\int du T_{uu}^{\rm CFT}$. This also translates the Gao-Wald theorem (which states `Bulk ANEC $\Rightarrow$ Causality') into CFT language (section \ref{ss:lightrayrel}).
\item We show that the Regge OPE of CFT probe operators satisfies an `equation of motion' in which it is sourced by a bulk light-ray operator,
\begin{align}\label{finalReggeCintro}
(\hat{C}_{12}-2d)\left( \frac{\psi(x_1)\psi(x_2)}{\langle \psi(x_1) \psi(x_2)\rangle}-1\right) &= 4\pi G_N \Delta_\psi (u_2-u_1) \int du T_{uu}^{\rm bulk}\ ,
\end{align}
where $\hat{C}_{12}$ is the conformal Casimir operator (section \ref{ss:casope}). See \eqref{finalReggeC} for the precise equation with all kinematics defined.
\item The right-hand side of \eqref{finalReggeCintro} is positive by the bulk ANEC. Plugging this relation into a 4-point function $H(z,\bar{z})$ implies that
\begin{align}\label{introCas4}
(\hat{C} - 2d) \mbox{Disc} \, H(z,\bz) \geq 0 \ ,
\end{align}
in CFTs with an Einstein-gravity dual. This is a CFT manifestation of the focusing inequality. For comparison, the chaos sign bound \cite{Maldacena:2015waa} or CEMZ constraints \cite{Camanho:2014apa} in this setup require $\mbox{Disc}\, H \leq 0$. The focusing-like inequality \eqref{introCas4} is stronger. (Section \ref{s:fourpoint}).
\item Finally, we show by direct evaluation that \eqref{introCas4} also holds in conformal Regge theory, including stringy corrections. This is nontrivial and somewhat of a surprise, because it can no longer be interpreted as the bulk ANEC; we do know whether it is an accident or a hint of something deeper. (Section \ref{s:conformalregge}.)
\end{itemize}

\section{Focusing and the $S$-matrix}

In this section we will review some aspects of causality in Einstein gravity in asymptotically flat spacetime, explain its relation to the focusing bound sketched in the introduction, and show that the CEMZ causality constraints \cite{Camanho:2014apa} on three-point couplings can all be derived from the ANEC. We also provide some evidence that a generalized focusing bound holds beyond spin-2 gravity.

\subsection{Causality and the ANEC}\label{flatANEC}

Consider a high-energy probe particle scattering from a weak gravitational potential in $D$ dimensions. The spacetime metric is
\begin{align}\label{dudvcoords}
ds^2 = -du dv + d\vec{x}^{\,2} + h_{\mu\nu} dx^\mu dx^\nu + O(h_{\mu\nu}^2)  \ ,
\end{align}
where $u=t-y$, $v=t+y$, and $\vec{x}$ labels the transverse directions $\mathbb{R}^{D-2}$. A high-energy probe travels on a geodesic that is approximately null, so we can take the initial momentum to be purely in the $u$-direction, $P^\alpha = 2E \delta^\alpha_u$. At first order in the metric perturbation, for null trajectories at fixed impact parameter, one can integrate $ds^2=-du dv+h_{uu} du^2=0$ to find the time delay of the probe,
\begin{align}
\Delta v(\vec{x}) = \int_{-\infty}^{\infty} du h_{uu}(u,\vec{x}) \ .
\end{align}
The time delay marks the arrival of the probe particle at future null infinity, as a function of the impact parameter  $\vec{x}$ (having placed the target at $\vec{x} = \vec{0}$). Causality requires\footnote{We assume $D >3$. In $D=4$ there are IR divergences so this requires an IR cutoff $|\vec{x}| < L_{IR}$. The time delay is defined such that $\Delta v(L_{IR}) = 0$.} 
\begin{align}\label{dvYY}
\Delta v(\vec{x}) \geq 0 \ .
\end{align}
In Einstein gravity, classical causality follows from the averaged null energy condition (ANEC) \cite{A_Borde_1987, PhysRevD.33.3526, Friedman:1993ty, PhysRevLett.61.1446, Graham:2007va, Gao:2000ga}.  This holds at the nonlinear level but we will only review the linearized version. Using the Einstein equation and the linearized Ricci tensor $R_{uu} = -\frac{1}{2} \eta^{\alpha\beta}\p_\alpha \p_\beta h_{uu} + u$-derivatives, the time delay satisfies the Poisson equation
\begin{align}\label{focusvt}
\dsq \Delta v= -16 \pi G_N \int_{-\infty}^{\infty}du T_{uu} \ .
\end{align}
Thus the averaged null energy is the source for time delay. 
And the ANEC, in Einstein gravity, implies 
\begin{align}\label{focusdeltav}
\dsq \Delta v \leq 0 \ .
\end{align}
We refer to this as the (classical) focusing bound for reasons explained shortly.

At the classical level, the ANEC is stronger than causality:  \eqref{focusdeltav} implies that the time delay is non-negative, but the converse is not true. To prove the first part of this statement, note that \eqref{focusvt} can be inverted to solve for $\Delta v(\vec{x})$. The transverse Green's function is negative definite, so it follows that $\Delta v \geq 0$.\footnote{In $D=4$, \eqref{focusdeltav} implies \eqref{dvYY} for any choice of $L_{IR}$ larger than the target transverse size. For example, consider a target localized at $u=0$ and with a finite transverse profile. The stress energy-momentum tensor can be modeled as $T_{uu} =|P_u| \delta(u) f(\vec{x})$  where $P_u$ is the momentum of the high-energy target and $f(\vec{x}) > 0$ has compact support for $|\vec{x}| \leq L$. The solution to Eq.~\eqref{focusvt} is then $\Delta v(\vec{x}) = -16\pi G_N  |P_u| \int d^2 y G(\vec{x},\vec{y})f(\vec{y})$ where $G(\vec{x},\vec{y}) = \frac{1}{2\pi}\log({|\vec{x} - \vec{y}|/L_{IR}})$ is the regularized Green function of the $\dsq$ operator with the IR cutoff $L_{IR}$. The positivity of the time delay is then ensured as long as $L<L_{IR}$.}

\subsection{Focusing}\label{ss:focusing}

The quantity $\dsq \Delta v$ has a natural interpretation: It is the expansion of the null congruence defined by a family of probe particles in the perturbed geometry.

A congruence is a continuous family of curves in a region of spacetime such that within the region, exactly one curve passes through each point.  For a congruence of null geodesics, the tangent vector field $n^\alpha$ has $n^2=0$ and satisfies the geodesic equation, $n^\alpha \nabla_\alpha n^\nu = 0$. The expansion is defined
\begin{align}
\theta = \nabla_\mu n^\mu = \frac{1}{2E} \nabla_\mu P^\mu \ ,
\end{align}
where $P^\alpha = 2E n^\alpha$ is the canonically normalized $D$-momentum. For an irrotational\footnote{A congruence is irrotational if and only if it is orthogonal to a family of hypersurfaces. We will consider null geodesic congruences which initially have $\nabla_\alpha n_\beta = 0$ and this implies that they are irrotational everywhere.} congruence, this can be related to the transverse area element $\sqrt{\sigma}$ of the null hypersurface generated by the congruence as 
\begin{align}
\theta = \frac{1}{\sqrt{\sigma}} \frac{d}{d\lambda} \sqrt{\sigma} \ , 
\end{align}
where $\lambda$ is the affine parameter defined by $n^\mu\p_\mu = \frac{d}{d\lambda}$.
The Raychaudhuri equation determines how this quantity evolves along the congruence:
\begin{align}
\frac{d\theta}{d\lambda} = - \frac{1}{D-2}\theta^2 - \sigma_{\alpha\beta}\sigma^{\alpha\beta} 
- R_{\alpha\beta} n^\alpha n^\beta
\end{align}
where $R_{\alpha\beta}$ is the Ricci tensor, and $\sigma_{\alpha\beta}$ is the shear, a geometric property of the congruence that is linear in $\nabla_\alpha n_\beta$. In Einstein gravity, we can rewrite the curvature in terms of the matter stress tensor, 
\begin{align}
\frac{d\theta}{d\lambda} = - \frac{1}{D-2}\theta^2 - \sigma_{\alpha\beta}\sigma^{\alpha\beta} 
- 8\pi G_N T_{\alpha\beta} n^\alpha n^\beta \ .
\end{align}
The quadratic terms are manifestly negative, so assuming the null energy condition, this implies
\begin{align}
\frac{d\theta}{d\lambda} \leq 0 \ .
\end{align}
This is the focusing theorem, and it underlies many fundamental results in general relativity including Hawking's area theorem and the singularity theorems. It describes how the geometry reacts to a positive energy distribution. It also describes how energetic probe particles scatter from a gravitational potential, since high-energy probes travel on (approximately) null geodesics. 

Let us expand at the linearized level about Minkowski spacetime and assume the probes are initially parallel, so that at early times $n^\alpha = \delta^{\alpha}_u$ and $\nabla_\alpha n_\beta = 0$. 
To first order in $\nabla_\alpha n_\beta$ the Raychaudhuri equation becomes
$
\frac{d\theta}{d\lambda} \approx -R_{\alpha\beta}n^\alpha n^\beta  
$.
Using the linearized curvature, and integrating with $\theta_{\rm initial} = 0$, we find the late-time expansion
\begin{align}\label{thetaV}
\theta_{\rm final} = \frac{1}{2}\dsq \Delta v \ .
\end{align}
Therefore the inequality $\dsq \Delta v \leq 0$ is a special case of focusing. We have essentially just repeated the derivation of Hawking's area theorem, applied to the hypersurface generated by the null geodesics that begin at $u=-\infty$, $v=v_0$. This hypersurface is the past causal horizon for the portion  of past null infinity $v>v_0$. Therefore by Hawking's area theorem it has non-increasing area,  $\theta \leq 0$. (Note that Hawking's theorem for past horizons states that the area cannot increase; the sign is opposite for the more common case of a future horizon.)

\subsection{Graviton sources and higher curvature corrections}\label{ss:cemz}
\begin{figure}
\centering
\begin{tikzpicture}[node distance=0.5 cm and 1.5cm]
\coordinate[] (e1) at (0,0);
\coordinate[] (aux1) at (3,-0.5);
\coordinate[] (e2) at (6,0);
\coordinate[] (aux2) at (3,-2.5);
\coordinate[label=left:$h$] (e3) at (0,-3);
\coordinate[label=right:$h$] (e4) at (6,-3);
\draw[particle] (e1) -- (aux1);
\draw[particle] (aux1) -- (e2);
\draw[graviton] (e3) -- (aux2);
\draw[graviton] (aux2) -- (e4);
\draw[graviton] (aux2) -- node[label=right:$h$] {} (aux1);
\end{tikzpicture}
\caption{
Graviton-mediated scattering of a scalar and graviton. The generalized ANEC \eqref{genanec} implies that higher curvature corrections to the 3-graviton vertex must be suppressed by the cutoff, reproducing the constraints of \cite{Camanho:2014apa}.
} \label{fig:gravitonscalar}
\end{figure}
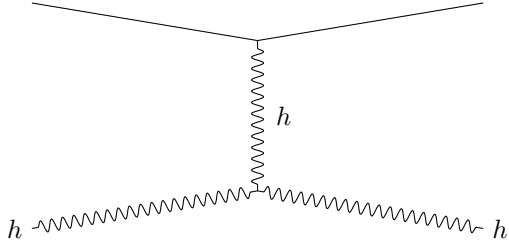

We will now extend the analysis to higher order in the metric perturbation and include higher curvature corrections. This covers all single-graviton-mediated effects on a scalar probe (with the scalar minimally coupled). 
In particular, it applies to the $2 \to 2$ scattering of a scalar and a graviton, in a theory with a general 3-graviton vertex; see figure \ref{fig:gravitonscalar}.\footnote{We work to $O(h_{\mu\nu}^2)$ in the sources, but $O(h_{\mu\nu})$ in the scalar-scalar-graviton interaction. This means we are keeping the diagram in Figure~\ref{fig:gravitonscalar}, but dropping the diagrams with an intermediate scalar or a $\phi^2 h_{\mu\nu}^2$ quartic interaction, which are subleading in the eikonal limit.}

The following equations still hold:
\begin{align}\label{stilltrue}
\theta_{\rm final} = \frac{1}{2} \dsq \Delta v = \frac{1}{2}\dsq \int_{-\infty}^{\infty} du  h_{uu}
\end{align}
These are kinematic relations satisfied by null geodesics in a perturbed background, so they do not depend on the equations of motion. The equations of motion for higher curvature gravity, expanded to second order, take the form

\begin{align}
G_{\mu\nu}^{(1)} = 8 \pi G_N T_{\mu\nu}^{\rm total} \ , 
\end{align}
where $G_{\mu\nu}^{(1)}$ is the linearized Einstein tensor. 
 $T_{\mu\nu}^{\rm total}$, which is defined by this equation, is the usual matter stress tensor $T_{\mu\nu}$ plus a pseudotensor quadratic in the metric field. The pseudotensor depends on the higher curvature couplings. Now we integrate over a null ray parameterized by $u$. Combined with \eqref{stilltrue}, this yields
\begin{align}\label{dvtg}
\dsq \Delta v = -16 \pi G_N \int_{-\infty}^{\infty} du T_{uu}^{\rm total} \ .
\end{align}
This generalizes the focusing equation to include graviton sources with arbitrary couplings. The equation \eqref{dvtg} can be inverted using the transverse Green's function,
\begin{align}
\Delta v(\vec{x}) = -\int du \int d\vec{x}' G(\vec{x}, \vec{x}') T_{uu}^{\rm total}(u,v=0,\vec{x}') \ .
\end{align}
 If the source on the right-hand side is localized in the transverse space, i.e. $T_{uu}^{\rm total} = t_{uu} \delta^{d-1}(\vec{x})$, then 
\begin{align}
\Delta v(\vec{x}) = - G(\vec{x},0) \int du\, t_{uu}(u,v=0) \ .
\end{align}
The Green's function is negative. Therefore, in such states, the following three conditions are equivalent:\footnote{Of course, `generalized ANEC $\Rightarrow$ causality' even for states that are delocalized. We needed to choose a state local in the transverse direction for the implication `causality $\Rightarrow$ generalized ANEC'.}
\begin{align}
\mbox{causality:}& \quad \Delta v \geq 0 \\
\mbox{focusing:}& \quad \dsq \Delta v \leq 0\\
\mbox{generalized ANEC:}& \quad \int du T_{uu}^{\rm total} \geq 0 \ .\label{genanec}
\end{align}
All three must be satisfied in consistent theories of a probe particle minimally coupled to higher derivative gravity. The third inequality \eqref{genanec} is the natural generalization of the ANEC to include contributions from the graviton. Shockwaves have local sources in the transverse space, so we conclude that the generalized ANEC \eqref{genanec} implies the constraints on graviton 3-point vertices derived in \cite{Camanho:2014apa} (CEMZ). That is, in a state $|\Psi\rangle$ defined by acting on the vacuum with the $h_{\mu\nu}(x)$ smeared against a wavepacket (as in \cite{Kosower:2018adc, Cristofoli:2020hnk}), the inequality $\langle \Psi | \int du T_{uu}^{\rm total}|\Psi\rangle \geq 0$ reproduces the CEMZ bounds.

\subsection{Example: Bounding higher derivatives operators with the ANEC}\label{ss:cemzexample}

Similarly, the ANEC reproduces the CEMZ constraints \cite{Camanho:2014apa} on other higher dimension operators. As an example, let us work with a gravitational EFT of a photon, whose leading Lagrangian is
\begin{equation}
\mathcal{L} = \frac{1}{4}F_{\mu\nu}F^{\mu\nu} + \frac{\hat{\alpha}_2}{4}R^{\mu\nu}_{\phantom{\mu\nu}\rho\sigma}F_{\mu\nu}F^{\rho\sigma}\,.
\end{equation} 
For such theory, the time delay experienced by a photon scattering off a shock-wave is \cite{Camanho:2014apa}
\begin{equation}
\label{timeDelayPhoton}
\Delta v(\vec{x})= \frac{4\Gamma\left(\frac{D-4}{2}\right)}{\pi^{\frac{D-4}{2}}}G_N|P_u|\left[1+\hat{\alpha}_2 \epsilon^i_h\epsilon^j_h\partial_{x^i}\partial_{x^j}\right]\frac{1}{|\vec{x}|^{D-4}}
\end{equation}
where $\epsilon^i_h$ is the polarization vector of the photon of helicity $h = \pm 1$ (normalized such that $\epsilon^2 = 1$) and  $P_u < 0$ is the momentum of the target generating the shockwave geometry. Eq.\eqref{timeDelayPhoton} can be also interpreted as the time delay experienced by a scalar probe travelling along a null geodesic in a shock-wave geometry generated by a non-minimally coupled photon of helicity $\epsilon_h$. If we apply the $\dsq$ operator to Eq.~\eqref{timeDelayPhoton}, we get
\begin{equation}
\dsq \Delta v(\vec{x}) = -16G_N|P_u|\left[1+\hat{\alpha}_2 \epsilon^i_h\epsilon^j_h\partial_{x^i}\partial_{x^j}\right]\delta^{D-2}(\vec{x})\,.
\end{equation}
This last equation should be understood in the sense of distributions. For example, if we scatter a wavepacket of scalar particles with profile $f(\vec{x})$, then Eq.~\eqref{focusdeltav} becomes
\begin{equation}
\label{focusingConvoluted}
\int d^{D-2}\vec{x} f(\vec{x})^2 \dsq \Delta v(\vec{x}) = -16G_N|P_u|\left[f(\vec{0})^2+\hat{\alpha}_2\, \epsilon^T_h\cdot H(\vec{0})\cdot \epsilon_h\right] \leq 0\, ,
\end{equation}
where  $H(\vec{x})_{ij} = \partial_{x^i}\partial_{x^j}f(\vec{x})^2$ is the Hessian matrix. For concreteness, we can consider a gaussian wavepacket $f(\vec{x}) = \text{exp}\left(-\frac{(\vec{x}-\vec{x}_*)^2}{4\sigma^2}\right)$ centered at some point $\vec{x}_*$. The width $\sigma$ is at least of the order of the precision at which we can localize the state within the EFT with physical cutoff $\Lambda$, that is $\sigma \sim 1/\Lambda$. For such a wavepacket, Eq.~\eqref{focusingConvoluted} becomes
\begin{equation}
\label{focusingConvoluted2}
\int d^{D-2}\vec{x} f(\vec{x})^2 \dsq \Delta v(\vec{x}) \propto -16G_N|P_u|\left[1+\frac{\hat{\alpha}_2}{\sigma^2}\left(\frac{\left(\vec{\epsilon}_h \cdot \vec{x}_*\right)^2}{\sigma^2}-1\right)\right] f(\vec{0})^2\leq 0\, ,
\end{equation}
which returns the two-sided bound
\begin{equation}
\label{boundalpha2}
|\hat{\alpha}_2| \lesssim \sigma^2 \sim \frac{1}{\Lambda^2}
\end{equation}
for $\vec{x}_* \simeq \vec{0}$ and $\vec{x}_* \simeq \sigma \vec{\epsilon}_h$. Note that higher-order corrections to Eq.\eqref{focusingConvoluted2} arising from higher-dimensional operators in the Lagrangian are $O(|\vec{x}_*|^{\,4}/\sigma^4)$ and therefore one cannot rely on bounds obtained for $|\vec{x}_*| \gg \sigma$. In fact, in this latter limit the wavepacket is exponentially suppressed around $\vec{x}\sim \vec{0}$ where $\dsq \Delta v(\vec{x})$ is localized.

The CEMZ bounds on higher curvature corrections to Einstein gravity are obtained by performing a very similar calculation. By the argument in section \ref{ss:cemz}, the photon-photon-graviton constraint derives from the ordinary ANEC, while the 3-graviton constraint comes from the  generalized ANEC which includes the pseudotensor (i.e., $T_{uu} \to T_{uu}^{\rm total}$).

\subsection{Relation to the $S$-matrix}\label{ss:relateS}
We will now rephrase the focusing bound in terms of the $S$-matrix. Let us consider a $2 \to 2$ scattering of scalars, and define the Mandelstam variables $s=-(p_1+p_2)^2, t= -(p_1-p_3)^2, u = -(p_1-p_4)^2$ where $p_i$ is the momentum of the $i-$th particle in the process. Given the scattering amplitude ${\cal M}(s,t)$, one can define the impact-parameter amplitude
\begin{align}
{\cal M}(s,\vec{x}) &= \int \frac{d^{D-2}\vec{q}}{(2\pi)^{D-2}} \, e^{i \vec{q}\cdot \vec{x}} {\cal M}(s,t=-\vec{q}^{\,2}) \,,
\end{align}
or, equivalently, the phase shift $\delta(s,\vec{x})$
\begin{align}
i{\cal M}(s,\vec{x}) =2s \left( e^{2i \delta(s,\vec{x})} - 1\right) \,. 
\end{align}
In the eikonal regime $s \gg 1/|\vec{x}|^2$, the amplitude exponentiates, so $\delta = \frac{1}{4s}{\cal M}_{\rm tree}$. 
By analyzing a wavepacket, the phase shift can be related to the time delay in the semiclassical limit,  
\begin{align}\label{dvreview}
\Delta v(\vec{x})  = 2\frac{\p}{\p E} \mbox{Re\ }\delta(s,\vec{x}) \ , 
\end{align}
where $E$ is the probe energy and $s=4E|P_u|$. Causality then requires
\begin{align}
\p_s \mbox{Re\ }\delta(s,\vec{x}) \gtrsim 0 \ .
\end{align}
The notation  `$\gtrsim$' means that it applies only to the high energy behavior, so in particular it only applies to a phase shift that grows as $s^{J-1}$ with $J>1$. Otherwise there is no semiclassical interpretation as a time delay. Acting with the transverse Laplacian we find
\begin{align}\label{dvdelta}
\dsq \Delta v = 2  \frac{\p s}{\p E} \frac{\p}{\p s} \dsq \mbox{Re\,} \delta(s, \vec{x}) \,. 
\end{align}

This can be related to the expansion of a congruence defined by a family of probes. The momentum transferred to a semiclassical probe particle is $\vec{Q} = 2  \mbox{Re\,} \vec{\partial} \delta(s,\vec{x})$ so a family of probes with initial momentum in the $u$-direction and final momentum $P^\alpha$ has expansion
\begin{align}\label{dpp}
\p_\mu P^\mu =2 \dsq\mbox{Re\,} \delta(s,\vec{x}) \ .
\end{align}
These are purely kinematic results that apply to any theory, and since Eq.\eqref{dpp} is evaluated at late times, it can always be interpreted as the expansion of a null geodesic congruence, even if the scattering was not via gravity.
In Einstein gravity, $\delta \propto s$, so Eq.\eqref{dvdelta} and Eq.\eqref{dpp} are consistent with the Raychaudhuri equation, which states $\p_\mu P^\mu = 2E \theta_f= E \dsq \Delta v$.

\subsection{Examples with higher spin fields}\label{ss:higherspin}

We will now discuss the extension of the focusing bound to UV completions of General Relativity, involving massive higher spin fields. The derivation above from the Raychaudhuri equation and ANEC does not apply to such theories. However, we will show by direct calculation that the focusing bound in the form
\begin{align}\label{realbound}
 \dsq \mbox{Re}\, \delta(s,\vec{x}) \leq 0 
\end{align}
also holds in string theory in the eikonal regime.

\subsubsection{Isolated higher spin particles}

Before we come to string theory in section \ref{venezianoSection}, it is instructive to discuss theories of isolated massive higher spin particles at weak coupling. These theories are known to be inconsistent with unitarity and causality even in flat spacetime
\cite{Velo:1969txo,Porrati:1993in,Cucchieri:1994tx,Camanho:2014apa, Arkani-Hamed:2017jhn, Afkhami-Jeddi:2018apj,Bellazzini:2019bzh}, and we will see that they are also incompatible with \eqref{realbound}.

Let us make some general remarks about the bound for real massless scalar amplitudes that satisfy the usual properties:
\begin{itemize}
\item \textbf{Analyticity:}  $\mathcal{M}(s,t)$ is analytic in the whole complex $s$-plane for fixed $t < 0$. The amplitude may have poles and branch-cuts on the real axis due to single and multi-particle production,
\item \textbf{Crossing symmetry:} For identical real scalar particles, crossing symmetry simply acts as $\mathcal{M}(s,t) = \mathcal{M}(t,s)= \mathcal{M}(u,t)$,
\item \textbf{Polynomially boundedness:} The amplitude at fixed $t$ is polynomially bounded at large $s$, i.e. $\mathcal{M}(s,t) < s^N$ for some integer $N>0$.
\end{itemize}
Under these assumptions, one can write a dispersive representation of the amplitude 
\begin{equation}
\label{disprel1}
\mathcal{M}(s,t) =I_\infty(s,t)+ \frac{1}{\pi}\int_{0}^{\infty}ds' \left[\frac{1}{s'-s}+\frac{1}{s'-u}\right]\text{Im}\mathcal{M}(s'+i \epsilon,t) 
\end{equation}
where $\text{Im}\mathcal{M}(s'+i \epsilon,t) \equiv \lim_{\epsilon\rightarrow 0^+}(\mathcal{M}(s'+i\epsilon,t)- \mathcal{M}(s'-i\epsilon,t))/2i$ and 
\begin{equation}
I_\infty(s,t)= \frac{1}{2\pi i }\oint_{C_{R\rightarrow\infty}} ds' \frac{\mathcal{M}(s',t)}{s'-s} 
\end{equation}
is an integral performed on a big circle $C_{R}$ on the complex plane with radius $R \rightarrow \infty$\footnote{Because of polynomially boundedness, $I_\infty(s,t)$ is just a polynomial in $s$ of order $N$. Indeed, $\partial_s^{N+1} I_\infty = 0$ and therefore $I_\infty(s,t) = \sum_{n=0}^{N} A_n(t)s^n$, where $A_{n}(t)$ are generic functions not fixed by analyticity. Note that divergences in $A_{n}(t)$ due to the contour integral at infinity are generically expected but they cancel out against the integral over the imaginary part. We then retain only the finite contributions in $I_\infty(s,t)$ and in the dispersive integral.}.
Using crossing symmetry $s \leftrightarrow t$, we can rewrite Eq.~\eqref{disprel1} as
\begin{equation}
\label{disprel2}
\mathcal{M}(s,t) =I_\infty(t,s)+ \frac{1}{\pi}\int_{0}^{\infty}ds' \left[\frac{1}{s'-t}+\frac{1}{s'-u}\right]\text{Im}\mathcal{M}(s'+i \epsilon,s)\,.
\end{equation}
For the purpose of computing the phase shift, in the eikonal limit $s\gg |t|$ only the first term of the integrand is relevant, returning a dispersive representation for the eikonal amplitude
\begin{equation}
\label{eikonalAmplitudeDisp}
\mathcal{M}_{eik}(s,t) = \frac{1}{\pi}\int_{0}^{\infty}ds' \frac{\text{Im}\mathcal{M}(s'+i \epsilon,s)}{s'-t}\,.
\end{equation}
The phase shift admits then the following dispersive representation
\begin{equation}
\label{phaseShiftDispersive}
\delta(s,\vec{x}) = \frac{1}{4\pi s}\int_0^\infty ds' \text{Im}\mathcal{M}(s'+i \epsilon,s)\int \frac{d^{D-2}\vec{q}}{(2\pi)^{D-2}} \, \frac{e^{i \vec{q}\cdot \vec{x}} }{s'+\vec{q}^2}\,,
\end{equation}
and, by acting with the operator $\dsq$, one gets 
\begin{equation}
\label{dispConjecture}
\dsq\delta(s,\vec{x}) = \int_0^\infty \frac{ds'}{4\pi s} \text{Im}\mathcal{M}(s'+i \epsilon,s)\left[-\delta^{(D-2)}(\vec{x}) + \int \frac{d^{D-2}\vec{q}}{(2\pi)^{D-2}} \, e^{i \vec{q}\cdot \vec{x}} \frac{s'}{s'+\vec{q}^2}\right]\,.
\end{equation}
Let us now consider a scalar-scalar scattering mediated by a tree-level exchange of a weakly coupled spin-$J$ particle of mass $m \neq 0$. In this case the imaginary part is just a delta function localized at the mass of the resonance, e.g.
\begin{equation}
\label{ImSingleParticle}
\text{Im}\mathcal{M}(s'+i \epsilon,s) = 16\pi^2 G s^J \delta(s'-m^2)
\end{equation}
where $G$ is some positive dimensionful constant depending on the underlying theory. The normalization of Eq.~\eqref{ImSingleParticle} is fixed such that, for graviton exchange in Einstein Gravity, $G$ is identified with the gravitational constant $G_N$ entering the Einstein equations. Plugging this expression back in Eq.~\eqref{dispConjecture}, we get 
\begin{equation}
\label{waveEqSingleParticle}
(\dsq-m^2)\delta(s,\vec{x}) = -2\pi G s^{J-1}\delta^{(D-2)}(\vec{x})\,,
\end{equation}
which shows that the focusing bound Eq.~\eqref{focusdeltav} is not satisfied because of the mass term. This differential equation just tells us that the phase shift is nothing else but the $D-2$ dimensional Fourier transform of a massive propagator,
\begin{equation}
\delta(s,\vec{x}) = \frac{G s^{J-1}}{\left(2\pi\right)^\alpha|\vec{x}|^{2\alpha}}  \left(|\vec{x}| m \right)^\alpha K_{\alpha}\left(|\vec{x}| m \right)
\end{equation}
where $\alpha = \frac{D-4}{2}$ and $K_{\alpha}\left(|\vec{x}| m \right)$ is the modified Bessel function. This equation shows explicitly that at weak coupling the phase shift is positive, even though the focusing bound Eq.~\eqref{focusdeltav} is not satisfied.

This conclusion applies to any finite sum of higher-spin exchanges. Conversely, in the next section, we will see that an infinite tower of higher-spins can indeed be consistent with the focusing bound.\footnote{ The importance of an infinite tower of states can be understood from the dispersive representation of the phase shift, Eq.~\eqref{phaseShiftDispersive}. For single particle exchanges, the phase-shift diverges at small impact parameters. This is not the case in string theory, where an infinite tower of massive higher spin states is exchanged, ensuring a smooth behavior at $\vec{x} = 0$. Eq.~\eqref{phaseShiftDispersive} allows us to understand a necessary (but not sufficient) condition for this transition to happen. At $\vec{x}=0$, the integral in the momenta $\vec{q}$ would naively diverge in the UV if $\text{Im}\mathcal{M}(s'+i \epsilon,s)$ is identically zero for some finite $s'$, that is if a finite tower of states is exchanged. For the phase shift to be a smooth function around $\vec{x}=0$, it is necessary to have an infinite tower of higher spin states. }

\subsubsection{Virasoro-Shapiro amplitude}
\label{venezianoSection}
String theory is another setting where we can explicitly test the focusing bound.
Let us consider the Virasoro-Shapiro amplitude for the scattering of 4 dilatons \cite{Virasoro:1969me}
\begin{equation}
\mathcal{M}(s,t,u) = -\frac{\pi G_N  \alpha'^3}{16}\frac{\Gamma\left(-\frac{\alpha's}{4}\right)\Gamma\left(-\frac{\alpha't}{4}\right)\Gamma\left(-\frac{\alpha'u}{4}\right)}{\Gamma\left(1+\frac{\alpha's}{4}\right)\Gamma\left(1+\frac{\alpha't}{4}\right)\Gamma\left(1+\frac{\alpha'u}{4}\right)}\left(s^4 + t^4 + u^4\right)
\end{equation}
where $\alpha'$ is the square of the string length.
In the high-energy limit $s\alpha' \gg 1 , s \gg |t|$, the amplitude exhibits the Regge behaviour 
\begin{equation}
\mathcal{M}(s,t,u) \sim -\frac{32 \pi G_N }{\alpha'}\frac{\Gamma\left(-\frac{\alpha't}{4}\right)}{\Gamma\left(1+\frac{\alpha't}{4}\right)}\left(-i\frac{s\alpha'}{4}\right)^{2\left(t\alpha'/4+1\right)}\,.
\end{equation}
Since $t<0$, in this limit the most important contribution comes from the regime of small $t$. As long as we take $\alpha'|t|\ll 1$, the ratio of gamma functions simplifies to a simple pole and we get the eikonal amplitude
\begin{equation}
\label{eikonalVeneziano}
\mathcal{M}(s,t,u) \sim \frac{128 \pi G_N}{\alpha'^2 t}\left(-i\frac{s\alpha'}{4}\right)^{2\left(t\alpha'/4+1\right)}\,.
\end{equation}
We are now ready to compute the derivative of the phase-shift Eq.~\eqref{focusdeltav}, 
\begin{align}
\dsq \delta(s,\vec{x}) &= -2\pi G_N s \int \frac{d^{D-2}\vec{q}}{(2\pi)^{D-2}} e^{i \vec{q}\cdot\vec{x}}e^{-2\vec{q}^{\,2}Y}\label{finalExpr_ddelta}
\end{align}
where $Y = \frac{\alpha'}{4}\log(-i\frac{s\alpha'}{4}) = \frac{\alpha'}{4}\left[\log(\frac{s\alpha'}{4})-i\pi/2\right]$. This is nothing else but the Fourier transform of a Gaussian distribution with a complex width,
\begin{equation}
\label{derivativeb2_phaseshift}
\dsq \delta(s,\vec{x})  =-\frac{2\pi G_N s}{(2\pi)^{D-2}}\times \left(\frac{\pi}{2Y}\right)^\frac{D-2}{2} e^{-\vec{x}^2/8Y}\,,
\end{equation}
with the real part
\begin{equation}
\label{ReLaplacianPhaseShit}
\text{Re}\,\dsq \delta(s,\vec{x}) = -\frac{2\pi G_N s}{(2\pi)^{D-2}}\left(\frac{\pi}{2|Y|}\right)^\frac{D-2}{2} e^{-\frac{\vec{x}^{\,2}}{2\alpha'\log\left(\frac{\alpha's}{4}\right)}}\cos\left({\frac{\vec{x}^{\, 2}\pi}{4\alpha' \log^2\left(\alpha' s/4\right)}}\right)\,.
\end{equation}
In the regime where this is reliable and not exponentially suppressed, i.e. $\alpha' \ll \vec{x}^{\,2} \ll \alpha'\log^2(s\alpha')$, it satisfies the focusing bound \eqref{realbound}.

\section{The CFT dual of the bulk ANEC}\label{s:anechkll}

We now turn to asymptotically anti-de Sitter spacetimes. Before we discuss focusing, we will answer an easier question: From a CFT point of view, why does QFT in AdS obey the ANEC?

In Poincare coordinates the AdS metric is 
\begin{equation}\label{poincareads}
    ds^2 = \frac{-dudv+dz^2+d\vec{x}^2}{z^2} \ .
\end{equation}
A path at fixed $z$, $\vec{x}$, $v$ is a null geodesic with affine parameter $u$. Define the averaged null energy operator by integrating over this geodesic,
\begin{equation}
{\cal E}_u^{\rm bulk} (z,v, \vec{x}) = \int_{-\infty}^{\infty} du T_{uu}^{\rm bulk}(z,u,v,\vec{x}) \ .
\end{equation}
In this section we consider a free scalar field in a fixed AdS background. Using the HKLL dictionary \cite{Hamilton:2005ju, Hamilton:2006fh, Hamilton:2006az}, we will rewrite 
${\cal E}_u^{\rm bulk}$ as a CFT operator in a form that is manifestly non-negative (see eq. \eqref{scalarEbulk}). Therefore the bulk QFT satisfies the ANEC,
\begin{align}
    \langle {\cal E}_u^{\rm bulk} \rangle \geq 0 \ .
\end{align}
The derivation can be viewed either as a bulk argument in a free QFT, or as a boundary argument in a generalized free CFT. The method is similar to the proof of the ANEC for a free scalar QFT in Minkowski spacetime \cite{Klinkhammer:1991ki}.

Let us emphasize that ${\cal E}_u^{\rm bulk}$ is the bulk ANEC operator, not the CFT ANEC operator which has been the subject of a large literature \cite{Hofman:2008ar, Kelly:2014mra, Hartman:2015lfa, Hofman:2016awc, 2017JHEP...07..066H, Kravchuk:2018htv}. The two are related by the Einstein equations, as we will describe in the next section.

\subsection{Derivation}

We work in $D=d+1$ bulk dimensions, in Poincare coordinates \eqref{poincareads} with $u=t-y,v=t+y$. We write $Y = (y,\vec{x}) \in \mathbb{R}^{d-1}$ for the boundary spatial coordinates, and $x = (t,y, \vec{x})$ for all the boundary coordinates.

The averaged null energy operator for a free scalar field is 
\begin{align}
    {\cal E}_u^{\rm bulk} = \int_{-\infty}^{\infty} du :\p_u  \phi \p_u \phi :\ .
\end{align}
The goal is to write this as a CFT operator.
The HKLL formalism \cite{Hamilton:2005ju, Hamilton:2006fh, Hamilton:2006az} tells us that the bulk scalar field $\phi$ can be reconstructed from the dual boundary operator ${\cal O}$ at leading order in $G_N$ as
\begin{align}\label{hkllphi}
    \phi(z,t,Y) &= \frac{\Gamma(\Delta-d/2+1)}{\pi^{d/2}\Gamma(\Delta-d+1)} \int_{t'^2 + Y'^2
     < z^2} dt' d^{d-1}Y' \left( \frac{z^2-t'^2 -Y'^2}{z}\right)^{\Delta-d} {\cal O}(t+t', Y + i Y')
\end{align}
with $\Delta = \frac{d}{2} + \sqrt{\frac{d^2}{4}+m^2 }$. We will also write the scaling dimension as $\Delta = \frac{d}{2}+\nu$. Since ${\cal O}$ is a generalized free field at this order we may expand it in creation and annihilation operators \cite{Kravchuk:2018htv},
\be\label{Oexpansion}
{\cal O}(t,y,\vec{x})=N_{\Delta}^{-1/2} \int_{p>0} \frac{d^d p}{(2\pi)^d} |p|^{\nu}(a^{\dagger}(p) e^{-i p \cdot x}+a(p) e^{i p \cdot x})
\ee
with $[a(p), a^{\dagger}(q)] = (2\pi)^d \delta^d(p-q)$. Here $p>0$ means that the momentum is in the future lightcone ($p^t > 0, p^2<0$), and $|p| =  \sqrt{(p^t)^2 - k^2}$ with $k$ the spatial momentum. The normalization factor $N_{\Delta}=\frac{2^{2\Delta-1}\pi^{\frac{d-2}{2}}}{(2\pi)^d} \Gamma(\Delta) \Gamma(\Delta-\frac{d-2}{2})$ is positive. Let us insert \eqref{Oexpansion} into \eqref{hkllphi}. The Fourier transform of the smearing kernel is
\begin{align}
\int_{t'^2+Y'^2<z^2} &dt' d^{d-1}Y' (z^2-t'^2-Y'^2)^{\nu-d/2} e^{\pm(i p^t t'+k \cdot Y')} 
=2^{\nu} \pi^{d/2}\frac{z^{\nu}}{|p|^{\nu}} \Gamma(\nu-d/2+1) J_{\nu}(z|p|)  \ .
\end{align}
Thus
\begin{align}
\phi(z,x)=z^{d/2}2^{\nu} \Gamma(\nu+1) N_{\Delta}^{-1/2}  \int_{p>0} \frac{d^d p}{(2\pi)^d}  J_{\nu}(z |p|)(a^{\dagger}(p) e^{-i p \cdot x}+a(p) e^{i p \cdot x})
\end{align}
which is the standard HKLL dictionary in momentum space \cite{Hamilton:2006az}. Canonically quantizing the bulk field in AdS also leads to the same formula.

Now we can reconstruct the averaged null energy operator, 
\begin{align}\footnotesize
    {\cal E}_u^{\rm bulk}&(z,v,\vec{x}) = 
    z^{d}2^{2\nu} \Gamma(\nu+1)^2 N_{\Delta}^{-1} \int_{-\infty}^{\infty} du \int_{p,q>0} \frac{d^d p}{(2\pi)^d}\frac{d^d q}{(2\pi)^d} J_{\nu}(z|p|) J_{\nu}(z|q|)\\
& \quad \times p_u q_u\left( -a^{\dagger}(p)a^{\dagger}(q) e^{-i (p+q) \cdot x}-a(p)a(q)e^{i (p+q) \cdot x}+ 2 a^{\dagger}(p)a(q) e^{-i (p-q) \cdot x} 
\right)\notag
\end{align}
Here $x = (u,v,z,\vec{x})$ and the integral is taken along the null geodesic at fixed $z,v,\vec{x}$. The first two terms drop out because the $u$-integral imposes $p_u + q_u = 0$, while the momentum integrals are restricted to the forward lightcone; therefore the only potential contribution comes from $p_u=q_u=0$ and here the integrand vanishes. The remaining term is manifestly nonnegative:
\begin{align}\label{scalarEbulk}
    {\cal E}_u^{\rm bulk} &= \frac{4\pi^{d/2+1}\Gamma(\nu+1)}{\Gamma(\frac{d}{2}+\nu)}z^d \int_{-\infty}^{\infty} du A_u(x)^\dagger A_u(x) 
\end{align}
with
\begin{align}
A_u(x) &=  \int_{q>0} \frac{d^dq}{(2\pi)^d} J_\nu(z|q|) q_u a(q)e^{iq \cdot x}  \ .
\end{align}
This can be viewed as either a bulk or CFT expression.
In the limit $z \to 0$, the operator localizes onto a null ray in the boundary, and becomes proportional to the CFT light-ray operator $\int du :\!\p_u{\cal O} \p_u {\cal O}\!:$. More generally, it is a linear combination of double trace operators in the CFT that does not localize onto a boundary null ray.

\section{The Einstein equation in the Regge OPE}

At this point we have identified a positive CFT operator ${\cal E}_u^{\rm bulk}$ in holographic CFTs. We proved it explicitly only for bulk scalars but we will assume that it is positive in general for the bulk matter QFT. The purpose of this section is to understand what this positivity means for CFT observables in Regge-like kinematics. We consider a holographic CFT with a weakly coupled Einstein-gravity dual.

\subsection{Bulk vs.~Boundary light-ray operators}\label{ss:lightrayrel}
Consider a small excitation of AdS, with metric perturbation $h_{\mu\nu}$. In Poincare coordinates,
\begin{equation}
    ds^2 = \frac{-dudv+dz^2+d\vec{x}^2}{z^2} + h_{\mu\nu} dx^\mu dx^\nu \ .
\end{equation}
In the vacuum, a path at fixed $z$, $\vec{x}$, $v$ is a null geodesic with affine parameter $u$. After turning on the metric perturbation, this path is deflected. The perturbed line element at $O(h_{\mu\nu})$ is $ -\frac{du dv}{z^2} + h_{uu} du^2=0$, so the time delay is
\begin{equation}
    \Delta v = z^2 \int_{-\infty}^{\infty}du h_{uu}
\end{equation}
with the integral taken over the original geodesic.

Boundary causality requires $\Delta v \geq 0$. The meaning of this inequality in the dual CFT has been studied extensively; see e.g. \cite{Cornalba:2008sp,Hofman:2008ar,Afkhami-Jeddi:2016ntf,Li:2017lmh,Kulaxizi:2017ixa, Meltzer:2017rtf,Costa:2017twz,Balakrishnan:2017bjg, Afkhami-Jeddi:2018own,Afkhami-Jeddi:2018apj,Kulaxizi:2018dxo,Kaplan:2019soo,Chandorkar:2021viw, Kundu:2021qpi,Caron-Huot:2022lff}. In the  limit $z \to 0$, the probe particle propagates near the boundary, and the dictionary is $h_{uu} \sim z^{d-2}T_{uu}^{\rm CFT}$, so $\Delta v \geq 0$ becomes equivalent to the CFT ANEC: $\int du \langle T_{uu}^{\rm CFT}\rangle \geq 0$ \cite{Hofman:2008ar, Kelly:2014mra, 2017JHEP...07..066H}. At finite $z$, $\Delta v \geq 0$ translates into non-negativity of the CFT length operator $L$ defined below  \cite{2017arXiv170903597A}. 

Here our main interest is in the bulk light-ray operator, $\int du  T_{uu}^{\rm bulk}$, whose expectation value is positive by the bulk ANEC. As in the flat-space discussion in section \ref{flatANEC}, this operator is related to the time delay by the Einstein equation. The linearized Einstein equation, integrated over $u$, is 
\begin{equation}\label{adsvt}
    [\Delta_H - (d-1)]\left( z \int_{-\infty}^{\infty} du h_{uu}\right)  = -16\pi G_N z \int_{-\infty}^{\infty} du \langle T_{uu}^{\rm bulk}\rangle \ ,
\end{equation}
where $\Delta_H = z^{d-1}\p_z (z^{-d+3} \p_z) + z^2 \vec{\p}^2 $ is the Laplacian in the transverse $(d-1)$-dimensional hyperbolic space $H_{d-1}$, with
\begin{equation}
    ds^2_H = \frac{dz^2 + d\vec{x}^2}{z^2} \ .
\end{equation}
Therefore, in terms of the time delay,
\begin{equation}\label{dvADS}
    \frac{1}{z}\left[ \Delta_H - (d-1)\right] \frac{\Delta v}{z} = -16 \pi G_N \int_{-\infty}^{\infty} du \langle T_{uu}^{\rm bulk} \rangle \ .
\end{equation}
The flat space limit of this equation corresponds to taking $\vec{\p}^2$ much larger than the other terms on the left, in which case we recover \eqref{focusvt}.

The equation \eqref{adsvt} can be inverted as
\begin{align}
\int_{-\infty}^{\infty} du h_{uu}(u,v,z,\vec{x}) &=-16\pi G_N \frac{1}{z} \int dz' d\vec{x}' \sqrt{g_H} z' G(z',\vec{x}'; z,\vec{x})\langle {\cal E}_{u}^{\rm bulk}(v,z',\vec{x}') \rangle
\end{align}
where $G$ is the transverse Green's function. This Green's function is negative everywhere, so if the bulk ANEC is satisfied, then the time delay is positive. This is a theorem of Gao and Wald \cite{Gao:2000ga}, and here we followed the derivation in \cite{Engelhardt:2016aoo}. Now if we take $z\to 0$ and use the dictionary $h_{uu} \sim z^{d-2}T_{uu}^{\rm CFT}$, we find 
\begin{align}\label{EErel}
{\cal E}_u^{\rm CFT}(v, \vec{x}) = -  16\pi G_N a \int dz' d\vec{x}' \sqrt{g_H} z' K(z', \vec{x}'; \vec{x}) {\cal E}_u^{\rm bulk}(v, z', \vec{x}') \ ,
\end{align}
where $a$ is a positive constant and $K$ is the bulk-to-boundary propagator in $H_{d-1}$, which is negative. 
We have removed the expectation values because this can be interpreted as an operator equation in the boundary CFT. It relates a boundary light-ray operator to a bulk light-ray operator and implies that
\begin{align}
\mbox{Bulk ANEC} \quad \Rightarrow \quad \mbox{Boundary ANEC} \ .
\end{align}
Thus \eqref{EErel} is a CFT counterpart of the Gao-Wald theorem. The converse would not be true: Positivity on the left-hand side of \eqref{EErel} does not imply positivity of the source on the right (making no further assumptions about the bulk QFT) \cite{Engelhardt:2016aoo}. In this sense, the bulk ANEC appears to be stronger than the boundary ANEC.

\subsection{Translation into the OPE}\label{ss:casope}
The classical time delay of a probe particle is related to the Regge OPE in the dual CFT \cite{2017arXiv170903597A}. Consider a primary scalar CFT operator $\psi$, with conformal weight $1 \ll \Delta_\psi \ll C_T$ so that it is dual to a massive probe particle.  The gravitational (plus identity) contribution to the OPE is
\begin{align}\label{psieL}
    \psi(x_1) \psi(x_2) &\approx  e^{-m L(x_1,x_2)} \ , 
\end{align}
where $m$ is the mass and $L$ is the length operator, whose expectation value is the regulated length of a geodesic through the bulk from $x_1$ to $x_2$. This formula applies to probe particles minimally coupled to gravity. Let us linearize around the AdS vacuum and denote the contribution of the bulk graviton to the boundary OPE by $\psi(x_1) \psi(x_2)|_h$. By linearizing \eqref{psieL} we find
\begin{align}\label{linope}
\psi(x_1) \psi(x_2)|_h &= \langle \psi(x_1) \psi(x_2)\rangle (-m \delta L) \ .
\end{align}
The operator $\delta L$ can be expressed in terms of the bulk metric perturbation $h_{\mu\nu}$ integrated over the original geodesic. 
In the Regge limit this takes a particularly nice form. The Regge limit is defined by performing a large relative boost on the points $x_1$, $x_2$. In the coordinates $(u,v,\vec{x})$, the limit is taken as
\begin{align}\label{reggeUV}
u_2 \to \infty , \quad u_1 \to -\infty , \quad v_2 \to 0^- , \quad v_1 \to 0^+ 
\end{align}
while holding fixed the combination
\begin{align}
\Delta u \Delta v \quad \mbox{with} \quad \Delta u = u_2-u_1 , \ \Delta v = v_2-v_1
\end{align}
as well as the transverse positions, $\vec{x}_{1,2}$.
We are discussing an OPE that will eventually be plugged into a correlation function, and it is assumed that at that stage all other operator insertions will be held fixed in some bounded region as the limit is taken.

In the limit \eqref{reggeUV}, the geodesic connecting the points $x_1$ and $x_2$ on the boundary becomes nearly null, and away from the endpoints, it sits at fixed radial position $z=z_0$ and transverse position $\vec{x} = \vec{x}_0$, with
\begin{align}
z_0^2 &= \frac{1}{4}|x_1-x_2|^2 = \frac{1}{4}(- \Delta u \Delta v + (\Delta \vec{x})^2 )\\
\vec{x}_0 &= \frac{1}{2}(\vec{x}_1+\vec{x}_2) \ .
\end{align}
The variation of the geodesic length simplifies to\footnote{This is derived in equation (3.7) of \cite{2017arXiv170903597A}. We have generalized that equation slightly by labeling the points $x_{1,2}$ separately and allowing for nonzero $\Delta \vec{x}$.}
\begin{align}
\delta L = \frac{u_2 - u_1}{4} \int_{u_1}^{u_2} du h_{uu}(u,v=0, z=z_0, \vec{x} = \vec{x}_0)\ .
\end{align}
Therefore the single-graviton contribution to the Regge OPE \eqref{linope} is
\begin{align}\label{psipsih}
\frac{\psi(x_1)\psi(x_2)|_h}{\langle \psi(x_1)\psi(x_2)\rangle}
&= - \frac{\Delta_\psi}{4}(u_2-u_1) \int_{u_1}^{u_2} du h_{uu}(u,v=0, z=z_0, \vec{x} = \vec{x}_0) \ .
\end{align}
This is for $\Delta_\psi \gg 1$; the Regge OPE for light operators is derived from Witten diagrams in appendix \ref{ss:lightreggeope}. 
In terms of CFT operators, this has two distinct contributions: The CFT stress tensor $T_{\mu\nu}^{\rm CFT}$, and multitrace operators, which we will see momentarily are related to $T_{\mu\nu}^{\rm bulk}$. That is,
\begin{align}
\psi(x_1)\psi(x_2)|_h = \psi(x_1)\psi(x_2)|_T + \psi(x_1)\psi(x_2)|_{\rm multitrace} \ ,
\end{align}
where the first term is the stress tensor OPE block (including the coefficient)\cite{Czech:2016xec}. 
The first term can be derived from \eqref{psipsih} by rewriting $h_{uu}$ in terms of $T_{\mu\nu}^{\rm CFT}$ smeared against the free HKLL kernel. We will not need the explicit expression but it can be found in \cite{2017arXiv170903597A}.

A convenient way to understand the multitrace term is to act with the conformal Casimir operator to remove the contribution of the stress tensor. Define the differential operator
\begin{align}
    \hat{C}_{12} = - \frac{1}{2}({\cal L}_1^{ab} + {\cal L}_2^{ab})({\cal L}_{1,ab} + {\cal L}_{2,ab})
\end{align}
where ${\cal L}_i^{ab}$ is the action of the conformal generator on $\psi(x_i)$, i.e., $[L_{ab}, \psi(x_i)] = {\cal L}_{i,ab}\psi(x_i)$. The stress tensor OPE block is an eigenfunction of the Casimir satisfying
\begin{align}\label{c12eigen}
    \left(\hat{C}_{12}  - 2d\right) \left.\psi(x_1)\psi(x_2)\right|_{T} = 0 \ .
\end{align}
The Casimir eigenvalue for a spin-$\ell$ field of dimension $\Delta$ is $\lambda_{\Delta,\ell} = \Delta(\Delta-d) + \ell(\ell+d-2)$, so for the stress tensor, the eigenvalue is $2d$. (See e.g. \cite{Simmons-Duffin:2016gjk} for a review.) Thus acting with $(\hat{C}_{12}-2d)$ on the OPE will leave behind only the multitraces. 

By a tedious but straightforward computation we find that acting with the Casimir differential operator on \eqref{psipsih} gives\footnote{The limits of integration have been taken to infinity, due to \eqref{reggeUV}. This is allowed only after acting with the Casimir operator.}
\begin{align}\label{c12aa}
\frac{(\hat{C}_{12}-2d)\psi(x_1)\psi(x_2)|_h}{\langle \psi(x_1) \psi(x_2)\rangle} &= -\frac{\Delta_\psi}{4z_0}(u_2-u_1) (\Delta_H - (d-1)) \left( z_0 \int_{-\infty}^{\infty} du h_{uu}(u, v=0, z=z_0, \vec{x} = \vec{x}_0)\right) \ ,
\end{align}
where $\Delta_H$ is the transverse hyperbolic Laplacian acting on $(z_0, \vec{x}_0)$. The right-hand side is of course the same operator that appears in the Einstein equation, as it must be, since this operator should annihilate the contribution of $T_{\mu\nu}^{\rm CFT}$. Up to the overall coefficient, this equation can be understood as follows. The graviton contribution to the OPE comes from the vertex diagram  

\begin{align} \label{vertexdiagram}
\Pi_{\mu\nu}(x_1, x_2; X_3) \quad &=\quad
			\vcenter{\hbox{\begin{tikzpicture}[scale=.7]
				\def\x{0.8}
				\draw[] (0,0) circle (4*\x cm);   
				\draw[] (-2.8*\x,2.8*\x) to (-1.5*\x,0);
				\draw[] (-2.8*\x,-2.8*\x) to (-1.5*\x,0);
    \draw[graviton] (-1.5*\x,0*\x) to (1.5*\x,0);
				\filldraw[] (-1.5*\x,0) circle (2pt);
				\filldraw[] (-2.8*\x,-2.8*\x)  circle (2pt);
				\filldraw[] (-2.8*\x,2.8*\x)  circle (2pt);
    \filldraw[] (1.5*\x,0*\x)  circle (2pt);			
				\node[anchor =east] at (-1.5*\x,0) {};  \node[anchor =east] at (2.8*\x,0) {$X_3$};
				\node[anchor=south] at (0,0) {$h$};	
				\node[anchor=north west] at (-4*\x,-2.8*\x) {$x_2$};
				\node[anchor=south west] at (-4*\x,2.8*\x) {$x_1$};
			\end{tikzpicture}}}
\end{align}
Invariance under the AdS isometries imposes 
\begin{align}
    \left({\cal L}_{1,ab} + {\cal L}_{2,ab} + {\cal L}^{\rm bulk}_{3,ab}\right) \Pi_{\mu \nu}(x_1, x_2; X_3) = 0 
\end{align}
where ${\cal L}^{\rm bulk}_{3,ab}$ is the Lie derivative along the AdS isometry corresponding to the conformal generator $L_{ab}$, acting on the bulk point $X_3$. Therefore the Casimir operator $\hat{C}_{12}$ can be pulled into the bulk,
\begin{align}
    \hat{C}_{12} \Pi_{\mu \nu}(x_1, x_2; X_3) = \hat{C}_3^{\rm bulk}  \Pi_{\mu \nu}(x_1, x_2; X_3)
\end{align}
where $\hat{C}_3^{\rm bulk}$ is the Casimir operator that acts on $X_3$ by the bulk isometries. This Casimir operator acts on the graviton propagator, so the eigenvalue equation \eqref{c12eigen} is proportional to the left-hand side of the Einstein equation, and this is exactly what we found in \eqref{c12aa}.

Now applying the Einstein equation \eqref{adsvt} to \eqref{c12aa}, we finally obtain
\begin{align}\label{finalReggeC}
\frac{(\hat{C}_{12}-2d)\psi(x_1)\psi(x_2)|_h}{\langle \psi(x_1) \psi(x_2)\rangle} &= 4\pi G_N \Delta_\psi (u_2-u_1) {\cal E}_u^{\rm bulk} (v=0, z=\frac{1}{2}|x_1-x_2|, \vec{x}=\frac{1}{2}(\vec{x}_1+\vec{x}_2))\ .
\end{align}

This equation, which was quoted in the introduction, describes how the bulk light-ray operator ${\cal E}_u^{\rm bulk} = \int du T_{uu}^{\rm bulk}$ sources the Regge OPE. Acting with the Casimir operator removes the stress tensor, and ${\cal E}_u^{\rm bulk}$ is coming from the multitrace contribuition. For a single free scalar field this operator was written explicitly in \eqref{scalarEbulk}, but more generally, it has contributions from all of the bulk fields. A similar equation for arbitrary $\Delta_\psi$ is derived in appendix \ref{appendix:a} by evaluating the Witten diagram without taking the heavy limit, with the result 
\begin{align}\label{caslight}
&\frac{(\hat{C}_{12}-2d)\psi(x_1)\psi(x_2)|_h}{\langle \psi(x_1) \psi(x_2)\rangle} \\
&=   \int d^{d-2}\vec{x}' dz' du'  \frac{ 
 8\pi G_N (u_2-u_1)(4z_0^2)^{\Delta_\psi}\Delta_\psi \Gamma(\Delta_\psi+\frac{1}{2}) z'^{2 \Delta_{\psi}-d+3}}{\pi^{\frac{d-1}{2}} \Gamma(\Delta_{\psi}-\frac{d}{2}+1) \Big((\vec{x}'-\vec{x}_0)^2+z_0^2+z'^2\Big)^{2\Delta_{\psi}+1}}  T_{uu}^{\rm bulk}(u',v'=0,z',\vec{x}') \notag
\end{align}
The integrand in \eqref{caslight} is a positive kernel times ${\cal E}_u^{\rm bulk}$. Thus for any $\Delta_\psi$, assuming the bulk ANEC, we immediately find a new inequality for the OPE. Let us write the OPE with the identity removed as
\begin{align}
\delta(\psi\psi) = \psi\psi - \langle \psi\psi\rangle \ .
\end{align}
Then \eqref{finalReggeC} implies that
\begin{align}\label{negativeOPE}
(\hat{C}_{12} - 2d)\delta(\psi(x_1)\psi(x_2)) \geq 0
\end{align}
in the Regge limit. This will be used to understand the consequences of the bulk ANEC inside correlation functions. The derivation of \eqref{negativeOPE} assumed a holographic CFT dual to Einstein gravity, though we will find some indications that it is true more generally.

\section{Consequences for CFT 4-point functions}\label{s:fourpoint}

The operator inequality \eqref{negativeOPE} means that the expectation value of the left-hand side is nonnegative in any state. In particular we can evaluate it in a state created by a scalar primary, ${\cal O}(x_3) |0\rangle$, to deduce an inequality for the 4-point functions. Denote the normalized 4-point function by
\begin{align}\label{defH}
    H(z,\bz) &= (z \bz)^{\Delta_\psi} G(z,\bz) = \frac{\langle {\cal O}(x_4) \psi(x_2) \psi(x_1) {\cal O}(x_3)\rangle}{\langle \psi(x_1)\psi(x_2)\rangle \langle {\cal O}(x_3){\cal O}(x_4)\rangle}
\end{align}
and the commutator as
\begin{align}
    \mbox{Disc}\, H(z,\bz) &= \frac{\langle {\cal O}(x_4) \psi(x_2) [\psi(x_1),\  {\cal O}(x_3)]\rangle}{\langle \psi(x_1)\psi(x_2)\rangle \langle {\cal O}(x_3){\cal O}(x_4)\rangle}
\end{align}
with the cross ratios $z$ and $\bz$ determined by
\bal
z \bar{z}  = \f{x_{12}^2 x_{34}^2}{x_{13}^2 x_{24}^2},\qquad (1-z)(1-\bz) = \f{x_{14}^2 x_{23}^2}{x_{13}^2 x_{24}^2} \ .
\eal
With this normalization, the OPE in the Euclidean limit $z,\bz \to 0$ is $H(z,\bz) = 1 + \cdots$. 
The Regge limit is \eqref{reggeUV} for points $x_1$, $x_2$, with $x_{3,4}$ held fixed, which corresponds to small cross ratios $|z|,|\bz|\to 0$ in the Lorentzian regime \cite{2007JHEP...12..005B, Cornalba:2007fs, Cornalba:2008qf, Costa:2012cb}. In reflection-positive  kinematics the cross ratios are purely imaginary:
\begin{align}\label{rpzz}
 z = i \lambda , \quad \bz = i \blambda , \quad
    \mbox{with} \quad
    0 < \lambda, \blambda \ll 1 \ .
\end{align}
All of the bounds stated below on $\mbox{Disc}\, H$ are understood to be in these kinematics. 
Since the Euclidean correlator is dominated by  identity in this limit, the discontinuity is simply
\begin{align}
\mbox{Disc}\, H(z,\bz) &\approx H(z,\bz) - 1  \ .
\end{align}
Let us denote the cross-ratios
\be
z=\sigma,\quad \bar{z}=\eta \sigma
\ee
so the Regge limit is $\sigma \to 0$ with $\eta$ fixed, and $\sigma = i\times$(positive) is reflection positive. In many interesting cases the Regge correlator behaves as \cite{Costa:2012cb}
\begin{align}\label{reggeform}
H(z,\bz) \approx  1 - f(\frac{\bz}{z}) z^{1-J}=1-f(\eta) \sigma^{1-J} \ ,
\end{align}
with $|f| \ll 1$. In these cases, unitarity and causality impose the `chaos bound' $J \leq 2$ \cite{Maldacena:2015waa}. Furthermore, assuming $J>1$, the phase of the correction must be such that $H \leq 1$, i.e.,
\begin{align}\label{discneg}
\mbox{Disc}\, H(z,\bz) \leq 0 \ ,
\end{align}
 in the reflection-positive kinematics \eqref{rpzz}. These bounds apply to two scenarios:
\begin{enumerate}
\item Large-$N$ CFTs in the Regge limit \cite{Maldacena:2015waa}. If the CFT is also holographic with an Einstein-gravity dual, then \eqref{discneg} is the positivity of the bulk length operator \cite{2017arXiv170903597A}.
\item Any CFT in the lightcone limit $|\bz| \ll |z| \ll 1$ \cite{Hartman:2015lfa}. In this case the dominant
contribution to Disc $H$ is proportional to (minus) the CFT averaged null energy, $-{\cal E}_u^{\rm CFT}$, so \eqref{discneg} implies the CFT ANEC \cite{2017JHEP...07..066H}. This is also a special case of the Lorenetzian inversion formula \cite{2017JHEP...09..078C}.
\end{enumerate}
The two scenarios are generally distinct, though they overlap for large-$N$ CFTs in the lightcone limit. 

Now we will study the consequences of the OPE inequality \eqref{negativeOPE}. 
The conformal Casimir operator acts in terms of the cross ratio as \cite{Dolan:2003hv}
\begin{align}
    \frac{\hat{C}_{12}
    \langle {\cal O}(x_4) \psi(x_2) \psi(x_1) {\cal O}(x_3)\rangle}{\langle \psi(x_1)\psi(x_2)\rangle \langle {\cal O}(x_3){\cal O}(x_4)\rangle}
    &= \hat{C}H(z,\bz)
\end{align}
with
\begin{align}
\hat{C} \equiv 2z^2 \pa_z (1-z) \pa_z + 2\bz^2 \pa_{\bz} (1-\bz) \pa_{\bz}+ 2(d-2) \f{z \bz}{z-\bz} \l[ (1-z) \pa_z - (1-\zb) \pa_{\zb} \r]
\end{align}
In the Regge limit, the important terms in this differential operator are
\begin{align}\label{ChatRegge}
    \hat{C}  &\approx 2z^2 \pa_z^2 + 2\bz^2 \pa_{\bz}^2 + 2(d-2) \f{z \bz}{z-\bz} \l(  \pa_z -  \pa_{\zb} \r)
\end{align}
Applying \eqref{negativeOPE} inside the 4-point function we find 
\begin{align}\label{casGbound}
    (\hat{C} -2d) \mbox{Disc}\ H(z,\bz) \geq 0 
\end{align}
This is the sought-after focusing inequality for the 4-point function in the reflection-positive Regge kinematics \eqref{rpzz}. The derivation assumed a holographic CFT with an Einstein gravity dual. As we have shown below \eqref{EErel}, the differential bound \eqref{casGbound} implies the ordinary bound \eqref{discneg} and is generally stronger.

A natural question is whether \eqref{casGbound} is the holographic limit of a more general inequality, universal to \textit{all} CFTs, similar to how the CFT ANEC is a universal consequence of \eqref{discneg}. A related question is whether it can be derived using dispersive sum rules \cite{Hartman:2015lfa, 2017JHEP...07..066H, 2017JHEP...09..078C, Carmi:2019cub,Kologlu:2019bco, Caron-Huot:2020adz, Caron-Huot:2021enk} (or other methods, e.g. \cite{Belin:2020lsr}). We do not know the answers to these questions, but we will see below in section \ref{s:conformalregge} that it continues to hold in some cases beyond Einstein gravity.

It is instructive to consider what \eqref{casGbound} means in terms of the conformal block expansion. Assuming only graviton exchange in the bulk, we can expand
\begin{align}
H(z,\bz) = 1 + c_{\psi\psi T}c_{OOT} g_{d,2}(z,\bz) + H_{\rm double\ trace} \ ,
\end{align}
where the second term is the stress tensor conformal block and the last term is an infinite sum of conformal blocks for double trace operators $[\psi\psi]$ and $[{\cal O} {\cal O}]$.\footnote{From a CFT point of view, the double trace operators are necessary to cure the unphysical $\eta \to 1$ singularity from the boundary stress tensor block contribution in the correlators\cite{Gary:2009ae,Maldacena:2015iua}.  From a bulk point of view, these operators encode the gravitational dressing at order $1/N^2$, necessary for the reconstruction of a well-defined local bulk operator from the boundary\cite{Kabat:2011rz, Kabat:2012hp, Kabat:2012av, Kabat:2013wga, Kabat:2015swa}. We will see below how these features fit nicely together for the bulk stress tensor in $D>3$ and the connection to the Einstein equations, generalizing the observation for bulk scalar fields and stress tensor in $D=3$ \cite{Kabat:2015swa, Kabat:2016zzr, Lewkowycz:2016ukf, Kabat:2020nvj}.} Taking the discontinuity removes the identity term, and acting with $\hat{C}-2d$ removes the stress tensor, so only the double traces survive:
\begin{align}
(\hat{C} - 2d)\mbox{Disc}\,  H(z,\bz) = (\hat{C} - 2d)H_{\rm double\ trace}  \ . 
\end{align} 
Therefore
\begin{align}\label{dtsource}
(\hat{C} - 2d)H_{\rm double\ trace} &= 4\pi G_N(u_2-u_1) \int d^{d-2} \vec{x}' dz' k(\vec{x}_0,z_0;\vec{x}',z')  \langle {\cal O}(x_4) {\cal E}_u^{\rm bulk}(v=0, z', \vec{x}') {\cal O}(x_3)\rangle
\end{align}
where $k$ is the positive kernel in \eqref{caslight}.

\subsubsection*{Comparison to the explicit 4-point function}
To be more explicit, let us set $d=4$. In the Regge limit, the Witten diagram for graviton exchange in the $\langle {\cal O} \psi \psi {\cal O}\rangle$ correlation function gives a correlator of the form \eqref{reggeform} with $J=2$ \cite{Costa:2012cb},
\begin{align}
\mbox{Disc}\, H=-\frac{f(\eta)}{\sigma}= -\frac{i}{\sigma\sqrt{\eta}} C_{\psi \psi T} C_{OOT} \int_{-\infty}^{\infty} d\nu \f{90\pi \Gamma(\Delta_\psi - \f{i \nu}{2}) \Gamma(\Delta_\psi + \f{i \nu}{2}) \Gamma(\Delta_O - \f{i \nu}{2}) \Gamma(\Delta_O+ \f{i \nu}{2})}{(\nu^2+4) \Gamma(\Delta_\psi-1)(\Gamma(\Delta_\psi+1) \Gamma(\Delta_O-1) \Gamma(\Delta_O+1)} \Omega_{i\nu} (\eta) 
\end{align}
where $C_{OOT} C_{\psi \psi T}=\frac{8}{45} \Delta_\psi \Delta_O G_N$. $\Omega_{i\nu}$ is the harmonic function on hyperbolic space $H_{d-1}$. 
In $d=4$, it has a simple expression
\bal
\Omega_{i\nu}(x)=\frac{i \nu  x^{\frac{1}{2}-\frac{i \nu }{2}}}{4 \pi ^2 (x-1)}-\frac{i \nu  x^{\frac{1}{2}+\frac{i \nu }{2}}}{4 \pi ^2 (x-1)}.
\eal
Acting with $\hat{C}-8$ on $\mbox{Disc}\, H$ in $d=4$ and taking the $\Delta_\psi \to \infty$ limit, we get
\be
(\hat{C}-8)\mbox{Disc}\, H=i 8 \Delta_\psi \Delta_O G_N \int_{-\infty}^{\infty} d\nu \frac{  \Gamma(\Delta_{O}-\frac{i \nu}{2})\Gamma(\Delta_{O}+\frac{i \nu}{2})}{ \Gamma(\Delta_{O}-1) \Gamma(\Delta_{O}+1)} \frac{i \nu \eta^{-\frac{i\nu}{2}}}{\pi(\eta-1) \sigma}
\ee
where we have used shadow symmetry $\nu \to -\nu$ to simplify the expression. This integral can be calculated by summing over all the residues corresponding to the double trace operators, and the answer is 
\be\label{explicit4}
(\hat{C}-8)\mbox{Disc}\, H=\frac{32 i  \Delta_\psi \Delta_O G_N \eta^{\Delta_{O}} \Gamma(2 \Delta_{O}+1)}{\sigma  \Gamma(\Delta_{O}-1) \Gamma(\Delta_{O}+1) (1+\eta)^{2\Delta_{O}+1}} 
\ee
which is positive in the regime \eqref{rpzz}. This is the contribution of the operator \eqref{scalarEbulk}.

\section{Focusing in conformal Regge theory}\label{s:conformalregge}

The conceptual origin and derivation of the focusing bound $(\hat{C}-2d)\mbox{Disc}\, H \geq 0$ assumed an interaction dominated by graviton exchange. Nonetheless, we will now show that this inequality continues to hold in conformal Regge theory beyond the gravity limit. We do not know whether this is an accident, or a deep fact about quantum gravity and CFT. 

\subsection{Regge limit} 
\begin{figure} \label{n=4}
\centering
\includegraphics[width=8cm]{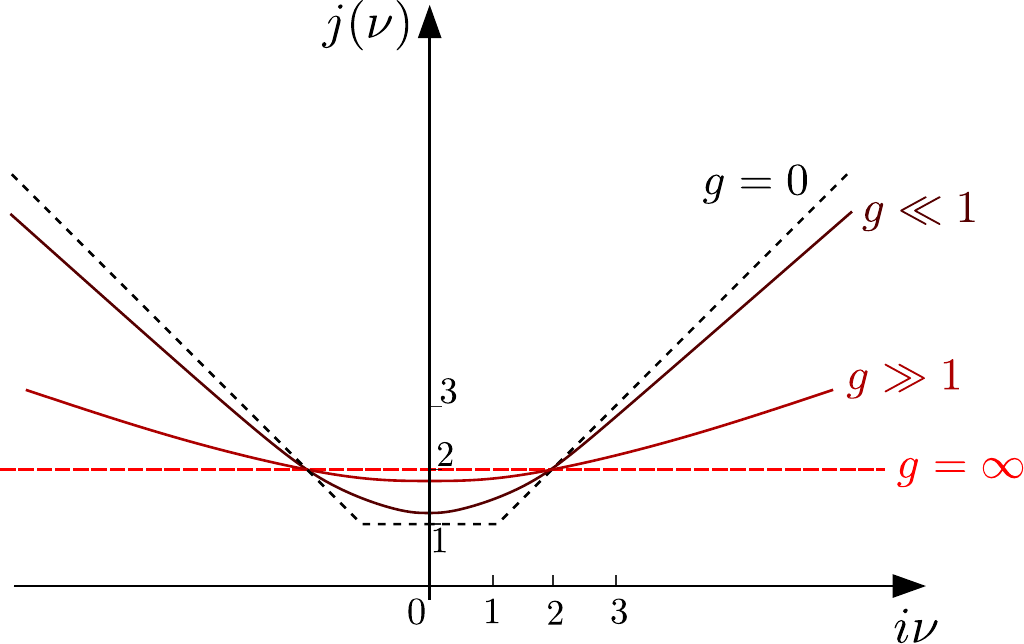}
\caption{The large-$N$ leading Regge trajectory for ${\cal N}=4$ SYM at different values of the coupling constant. }
\label{fig:birds}
\end{figure}

As in section \ref{s:fourpoint} we use the notation
\begin{align}
H(z,\bz) = (z \bz)^{\Delta_\psi } \langle \psi(0) \psi(z,\bz) {\cal O}(1) {\cal O}(\infty)\rangle
\end{align}
and 
\begin{align}
z = \sigma , \quad \bz = \eta \sigma \ . 
\end{align}
The Regge limit is $\sigma \to 0$ with $\eta >0$ fixed, with Lorentzian kinematics and operators ordered as in \eqref{defH}. The prediction of conformal Regge theory, assuming the correlator is dominated by a single Regge pole from the leading trajectory, is \cite{Cornalba:2007fs, Cornalba:2008qf, Costa:2012cb}
\begin{align}\label{creggeH}
H(\eta,\sigma) = 1 + \mbox{Disc}\, H(\eta,\sigma) = 1+2 \pi i \int_{-\infty}^{\infty} d\nu \alpha(\nu) (\sigma\sqrt{\eta})^{1-j(\nu)}\Omega_{i\nu}(\eta)
\end{align}
The function $j(\nu)$ encodes the leading Regge trajectory $\Delta(j)$ with $\Delta = \frac{d}{2}+i\nu$. The existence of the stress tensor imposes $j(\pm i \frac{d}{2})=2$. The leading Regge trajectory for ${\cal N}=4$ SYM in the large $N$ limit is sketched for various values of the coupling constant in Figure~\ref{fig:birds} \cite{Costa:2012cb}. The meromorphic function $\alpha(\nu)$ has poles corresponding to physical operators. $\Omega_{i\nu}$ is the harmonic function in $H_{d-1}$, 
\be \label{harmonic}
\Omega_{i\nu}(\eta)=\frac{\nu  \sinh (\pi  \nu ) \Gamma \left(\frac{d}{2}+i \nu -1\right) \Gamma \left(\frac{d}{2}-i \nu -1\right) }{2^{d-1} \pi ^{\frac{d}{2}+\frac{1}{2}} \Gamma \left(\frac{d}{2}-\frac{1}{2}\right)} \, _2F_1\left(\frac{d}{2}+i \nu -1,\frac{d}{2}-i \nu -1;\frac{d}{2}-\frac{1}{2};-\left(\frac{\eta^{\frac{1}{4}}-\eta^{-\frac{1}{4}}}{2} \right)^2\right)
\ee
Let's act on $\mbox{Disc}\, H$ with $({\hat C}-2d)$. The harmonic function satisfies
\begin{align}
\hat{C} \left( (\sigma\sqrt{\eta})^{1-l}
 \Omega_{i \nu}(\eta) \right) = \lambda_{\frac{d}{2}\pm i \nu,l} \left( (\sigma\sqrt{\eta})^{1-l}
 \Omega_{i \nu}(\eta) \right) =  \left(-\nu ^2-\frac{d^2}{4}+l (l+d-2)\right) \left( (\sigma\sqrt{\eta})^{1-l}
 \Omega_{i \nu}(\eta) \right) \ .
\end{align}
Therefore
\begin{align}
(\hat{C}-2d)\mbox{Disc}\,H &= 2\pi i \int_{-\infty}^{\infty} d\nu \alpha(\nu) (\sigma \sqrt{\eta})^{1-j(\nu)}\Omega_{i\nu}(\eta)\left( j(\nu)( j(\nu)+d-2) - \nu^2 - \frac{d(d+8)}{4}\right) \ .
\end{align}
By design, the factor in parentheses cancels the graviton pole at $\nu = \pm i \frac{d}{2}$. If the CFT is dual to Einstein gravity, then the dominant contribution after removing the graviton pole is the bulk ANEC, and it is positive. But we can now push the calculation into the stringy regime. Let us suppose that, like in ${\cal N}=4$ SYM, the integral is dominated by a saddlepoint at $\nu = 0$. Then the extra factor from acting with $(\hat{C}-2d)$ does not affect the location of the saddlepoint, so we find
\begin{align}\label{rjj}
(\hat{C}-2d)\mbox{Disc}\,H = \left[j(0)(j(0)+d-2)-\frac{d(d+8)}{4}\right]\mbox{Disc}\,H \ .
\end{align}
In reflection-positive kinematics, $\mbox{Disc}\, H $ is negative by the chaos sign bound, so we see that the sign of $(\hat{C}-2d)\mbox{Disc}\,H$ is controlled by the Regge intercept, $j(0)$. The chaos growth bound requires $j(0) \leq 2$. For $j(0) \in (0,2)$ and any $d$ the prefactor is negative and therefore 
\begin{align}
(\hat{C}-2d)\mbox{Disc}\,H \geq 0 \ .
\end{align}
This agrees with the focusing inequality we derived for Einstein gravity.\footnote{In 2D gravity the bulk ANEC is related to the size operator in the SYK model \cite{Susskind:2018tei, Brown:2018kvn, Lin:2019qwu}, it would be interesting to understand how the bulk ANEC in $D>3$ is related the size operator, and also interpret the inequality in conformal Regge theory as a constraint on the size operator for theories that don't saturate the chaos bound \cite{Shenker:2014cwa, Gu:2021xaj}.}

\subsection{Velocity-dependent Regge limit}
We can also take a slightly different Regge limit, tuning both $\eta$ and $\sigma$ at the same time \cite{2017JHEP...10..197C, 2018JHEP...06..121K, Mezei:2019dfv}. This allows us to interpolate between the non-local stringy Regge regime $|\sigma| \ll \eta$ and the lightcone limit $\eta \ll |\sigma|$, where any CFT is dominated by the stress tensor (assuming no low-twist scalars). It is convenient to parameterize this limit by coordinates $x,T$ with
\begin{align}
\sigma = i e^{x-T} , \quad \eta = e^{-2x} \ .
\end{align}
These $(x,T)$ correspond to spatial position and time in a thermal OTOC \cite{Maldacena:2015waa}. The limit $T \to \infty$ at fixed $x$ is the standard Regge limit, discussed in the previous subsection. Alternatively, following \cite{Mezei:2019dfv}, we can set $x = vT$, for fixed $v$, and then take $T \to \infty$. We refer to this as the velocity dependent Regge limit. There is a corresponding velocity-dependent Lyapunov exponent that obeys the bound on chaos \cite{
2017JHEP...10..197C, 2018JHEP...06..121K, 2018PhRvB..98n4304K, 2019NatPh..16..199X,  Mezei:2019dfv}.

\begin{figure} \label{saddle}
\centering
\includegraphics[scale=0.7]{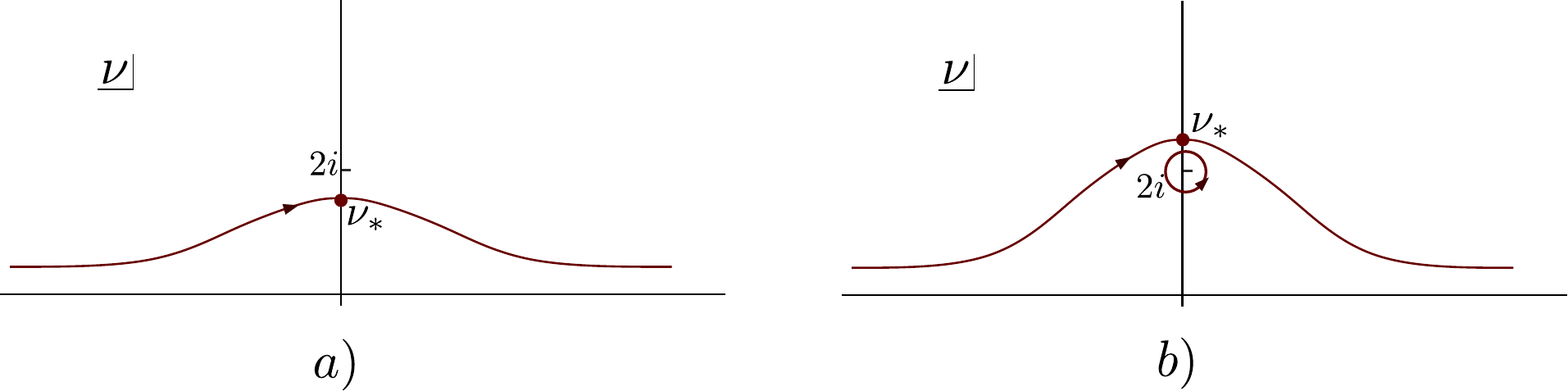}
\caption{a) When $v<4\mathcal{D} $, the integral in conformal Regge theory will be dominated by a saddle at $\nu=\nu_*$. b) When $v>4\mathcal{D} $, we also need to include the contribution from the stress tensor pole, and any other poles that are crossed in deforming the contour to through the saddle.}
\label{fig:saddlepole}
\end{figure}

The analysis that follows holds in general dimensions, but to be concrete we will specialize to $d=4$, where the harmonic function is
\begin{align}
\Omega_{i\nu}(\eta)=\frac{i \nu  \eta^{\frac{1}{2}-\frac{i \nu }{2}}}{4 \pi ^2 (\eta-1)}-\frac{i \nu  \eta^{\frac{1}{2}+\frac{i \nu }{2}}}{4 \pi ^2 (\eta-1)}
\end{align}
For $v>0$, we can drop the $\eta$ in the denominator. 
The conformal Regge amplitude, re-expressed in terms of $(x,T)$ and using the symmetry under $\nu \to -\nu$, is then
\begin{align}\label{vreg}
\mbox{Disc}\,H = \frac{1}{\pi} \int_{-\infty}^{\infty}d\nu \alpha(\nu) e^{i\pi(1-j(\nu))/2} e^{(j(\nu)-1-v+i\nu v)T}
\end{align}
To proceed, let us assume temporarily that $j(\nu)$ takes the same form as ${\cal N}=4$ SYM at strong coupling and large $N$, which is \cite{Cornalba:2007fs, 2007JHEP...12..005B}
\begin{align}\label{jlargen}
j(\nu) \approx  2 - \mathcal{D}(4+\nu^2) 
\end{align}
with $0< \mathcal{D} \ll 1$. 
We can now analyze the behavior of the integrand in \eqref{vreg} on the complex $\nu$-plane. The function $\alpha(\nu)$ has a pole at $\nu =  2i$, corresponding to the stress tensor, poles from double trace operators, and poles from higher spin physical operators along the imaginary axis. There is also a saddlepoint at
\begin{align}
\nu_* = \frac{iv}{2\mathcal{D}} \ .
\end{align}
The behavior of the integral depends on whether the saddle or the pole is closer to the real axis. This is illustrated in Figure~\ref{fig:saddlepole}.

If $v< 4\mathcal{D} $, then the contour of integration can be smoothly deformed to pass through the saddlepoint. Therefore we obtain
\begin{align}
(\hat{C}-2d)\mbox{Disc}\,H &= (\lambda_* - \lambda_T)\mbox{Disc}\,H 
\end{align}
where $\lambda_T=2d$ is the Casimir eigenvalue of the stress tensor, and $\lambda_* = \lambda_{\frac{d}{2}+i\nu_*, j(\nu_*)}$ is the eigenvalue of the saddlepoint. Using \eqref{jlargen}, this can also be written in $d=4$ as
\begin{align}\label{saddlecontrib}
(\hat{C}-8)\mbox{Disc}\,H
&= (j(\nu_*)-2)(j(\nu_*)+4+\frac{1}{\mathcal{D} })\mbox{Disc}\,H \ .
\end{align}
Now, choosing reflection positive kinematics, the chaos sign bound $\mbox{Disc}\, H < 0$ together with the chaos growth bound $j(\nu_*) < 2$ imply that the focusing bound is satisfied: $(\hat{C}-2d)\mbox{Disc}\,H \geq 0$. This inequality is sharp, in the sense that as $v \to 4\mathcal{D} $ from below, $j(\nu_*) \to 2$ and the focusing bound is saturated.

If the velocity $v$ is slightly above $4\mathcal{D} $, then when we deform the contour on the $\nu$ plane, we will pick up a contribution from the stress tensor pole, which now dominates $\mbox{Disc}\, H$. Therefore $(\hat{C}-2d)$ acting on the leading contribution vanishes. There is a subleading contribution to $(\hat{C}-2d)\mbox{Disc}\, H$ from the saddle, but at least in a neighborhood of $v \sim 4\mathcal{D} $, it remains positive --- the prefactor in \eqref{saddlecontrib} changes sign because of the factor $(j(\nu_*)-2)$, but $\mbox{Disc}\, H$ also changes sign because we've crossed the stress tensor pole in $\alpha(\nu)$ and there is a factor of $\frac{1}{\nu^2 + d^2/4}$.

As $v$ is increased even further to $v>4 \mathcal{D}  \Delta_O$, it will cross double trace poles, which now must be added to the saddlepoint contribution. The spin-2 double traces dominate, and for any finite $v$, we have $\eta \ll 1$ so it is always the lowest dimension spin-2 double trace that determines $(\hat{C}-2d)\mbox{Disc}\, H$. This contribution is equal to the leading term in \eqref{explicit4} as $\eta \to 0$. It comes from the bulk ANEC, so the focusing inequality still holds.

\ \\
\bigskip

\noindent \textbf{Acknowledgments} We thank Nima Afkhami-Jeddi, Brando Bellazzini, Yingfei Gu, Gary Horowitz, Giulia Isabella, Sandipan Kundu, Don Marolf, Gregoire Mathys, David Meltzer, Mark Mezei, Eric Perlmutter,  Jiaxin Qiao,  Andy Strominger, and Sasha Zhiboedov for helpful discussions. The work of T.H. and Y.J. is supported by the Simons Foundation through the Simons Collaboration on the Nonperturbative Bootstrap.  
The work of A.T. was supported in part by a grant from the Simons foundation and in part by funds from the University of California. The work of T.H., Y.J. and F.S. is supported in part by the NSF grant PHY-2014071. F.S. is also supported by a Klarman Fellowship at Cornell University.

\appendix

\section{Bulk ANEC operator in the Regge OPE}
\label{appendix:a}
In section \ref{ss:casope} we derived the contribution of the bulk averaged null energy operator, ${\cal E}_u^{\rm bulk}$, to the Regge OPE of $\psi\psi$ when $\Delta_\psi \gg 1$. In this appendix we use the Witten diagrams to derive the analogous equation \eqref{caslight} when $\psi$ is a light operator.

\subsection{Feynman rules}

We first summarize the Feynman rules to be used in the Witten diagrams \cite{DHoker:1999kzh, 2017arXiv170903597A}. We use Poincare coordinates \eqref{poincareads} with $u=t-y,v=t+y$. 
The scalar bulk-to-boundary propagator between a bulk point $(z,x)$ and a boundary point $x'$ is 
\be
D(z,x;x')=c_\Delta \frac{z^\Delta}{(z^2+(x-x')^2)^\Delta}\ ,
\ee
where
\be
c_\Delta=\frac{\Gamma(\Delta)}{\pi^{d/2}\Gamma(\Delta-d/2)}\ .
\ee
$\Delta=\frac{1}{2}(d+\sqrt{d+4m^2})$ is the scaling dimension of the boundary operator which is dual to a bulk scalar field of mass $m$, and  $(x-x')^2$ is the Minkowski (or Euclidean) norm in $d$ dimensions.

The bulk stress tensor vertex function is
\be\label{gss}\small
T^{\psi,\rm bulk}_{\mu\nu}(z,x;x_1,x_2)=T^{\rm bulk}_{\mu\nu}(D^\psi_1;D^\psi_2)={\cal N}_\psi \left(\partial_\mu D^\psi_1 \partial_\nu D^\psi_2+\partial_\nu D^\psi_1 \partial_\mu D^\psi_2 -(\partial_\alpha D^\psi_1)(\partial^\alpha D^\psi_2) g_{\mu\nu}- m^2 D^\psi_1 D^\psi_2 g_{\mu\nu}\right)\ .
\ee
where
\be
D^\psi_1\equiv D^\psi(z,x;x_1)\ , \qquad D^\psi_2\equiv D^\psi(z,x;x_2)\ 
\ee
and 
\be\label{defN}
{\cal N}_\psi=\frac{1}{ c_{\Delta_\psi} (2\Delta_\psi-d)}\ .
\ee
The derivatives in (\ref{gss}) are taken with respect to the bulk point $(z,x)$.
For the graviton propagator, the relevant components in the Regge limit \eqref{reggeUV} are \cite{DHoker:1999kzh, 2017arXiv170903597A} 
\be
G_{uu \alpha \beta }=(8\pi G_N) 2\partial_u \partial_\alpha \mathcal{U} \partial_u \partial_\beta \mathcal{U} G( \mathcal{U})
\ee
where 
\be
\mathcal{U}=-1+\frac{1}{\zeta},\quad \zeta=\frac{2 z_1 z_2}{z_1^2+z_2^2+(y_1-y_2)^2}
\ee
We have
\be
\partial_u \partial_v \mathcal{U}=\frac{1}{2 z_1 z_2} \ .
\ee
$G( \mathcal{U})$ is the bulk to bulk propagator for a massless scalar field, which satisfies
\be
\Delta G=\frac{i}{\sqrt{-g}} \delta^{d+1}(x'-x)=2 i z'^{d+1} \delta (u'-u) \delta (v'-v) \delta^2 (\vec{x}'-\vec{x}) \delta(z'-z)
\ee

\subsection{Regge OPE for $\delta(\psi(x_2)\psi(x_1))$}\label{ss:lightreggeope}

The graviton-exchange Witten diagram is equal to
\be \label{eq26}
\langle O(x_4) \psi(x_2) \psi(x_1) O(x_3) \rangle|_{h}=i \int d^{d+1} x  \sqrt{-g} \Pi_{\mu \nu}(x_1,x_2;X_3)  T^{\mu \nu, O,\rm bulk}(z,x;x_3,x_4)
\ee
where the vertex $\Pi_{\mu \nu}$ is  
\be
\Pi_{\mu \nu}(x_1,x_2;X_3)=i \int d^{d+1}x' \sqrt{-g'} G^{\alpha \beta}_{\mu \nu}(z',x'; X_3) T^{{\psi, \rm bulk}}_{\alpha\beta}(z',x';x_1;x_2)\ .
\ee
See the figure in \ref{vertexdiagram}, where $X_3$ labels the bulk point $(z,x)$. In the Regge limit \eqref{reggeUV}, the dominant contribution is the $T_{vv}^{\psi, \rm bulk}$ component, and the Witten diagram becomes
\be\small
\Pi_{\mu \nu}(x_1,x_2;X_3)=i \int d^{d+1}x' \sqrt{-g'} G^{vv}_{\mu \nu}(z',x'; X_3) T^{{\psi, \rm bulk}}_{vv}(z',x';x_1,x_2) =i \int d^{d+1}x' \frac{1}{2 z'^{d+1}} 4 z'^4 G_{uu \mu \nu} T_{vv}^{\psi,\rm bulk}\ .
\ee
By definition, we have
\be\small
T_{vv}^{\psi,\rm bulk}=\frac{\left(\Gamma \left(-\frac{d}{2}+\Delta _{\psi }+1\right) \right)^{-1} \pi ^{-\frac{d}{2}} \Delta _{\psi } \left(u'-u_1\right) \left(u'-u_2 \right) \Gamma \left(\Delta _{\psi }+1\right) z'^{2 \Delta _{\psi }}}{ \left(-\left(u'-u_1\right) \left(v'-v_1 \right)+\left(\vec{x}'-\vec{x}_1\right){}^2+z'^2\right){}^{\Delta _{\psi }+1} \left(-\left(u'-u_2 \right) \left(v'-v_2 \right)+\left(\vec{x}'-\vec{x}_2\right){}^2+z'^2\right){}^{\Delta _{\psi }+1}}
\ee
This becomes a delta function in $v'$ when we take the Regge limit. The proportionality factor is fixed by integrating both sides with respect to $v'$, and we get
\be \label{reggevertex}
T_{vv}^{\psi,\rm bulk}=i \frac{u_2-u_1}{2} \frac{ \Delta_\psi \Gamma(\Delta_\psi+\frac{1}{2}) z'^{2 \Delta_{\psi}}}{\pi^{\frac{d-1}{2}} \Gamma(\Delta_{\psi}-\frac{d}{2}+1) \Big((\vec{x}'-\vec{x}_0)^2+z_0^2+z'^2\Big)^{2\Delta_{\psi}+1}}  \delta(v')
\ee
where
\begin{align}
z_0^2 &= \frac{1}{4}|x_1-x_2|^2 = \frac{1}{4}(- \Delta u \Delta v + (\Delta \vec{x})^2 )\\
\vec{x}_0 &= \frac{1}{2}(\vec{x}_1+\vec{x}_2) \ .
\end{align}
Thus in the Regge limit, we have
\be\footnotesize
\Pi_{\mu \nu}=-(u_2-u_1)\int d^{d-2}\vec{x}' dz' du'  \frac{\Delta_{\psi} \Gamma(\Delta_\psi+1/2) z'^{2 \Delta_{\psi}-d+3}}{\pi^{\frac{d-1}{2}} \Gamma(\Delta_{\psi}-\frac{d}{2}+1) \Big((\vec{x}'-\vec{x}_0)^2+z_0^2+z'^2\Big)^{2\Delta_{\psi}+1}} G_{uu \mu \nu}(u',v'=0,\vec{x}',z';z,x)
\ee
This implies the general OPE
\be\footnotesize\label{lightreggeOPE}
\delta(\psi(x_2) \psi(x_1))=-\frac{(u_2-u_1)}{2}\int d^{d-2}\vec{x}' dz' du'  \frac{\Delta_{\psi} \Gamma(\Delta_\psi+1/2) z'^{2 \Delta_{\psi}-d+3}}{\pi^{\frac{d-1}{2}} \Gamma(\Delta_{\psi}-\frac{d}{2}+1) \Big((\vec{x}'-\vec{x}_0)^2+z_0^2+z'^2\Big)^{2\Delta_{\psi}+1}} h_{uu}(u',v'=0,\vec{x}',z')
\ee
where $h_{uu}$ is the bulk metric operator. 
We can also check the heavy limit of this OPE, using the identity 
\be \label{heavylimit}
\lim_{\Delta_\psi \to \infty} \frac{\Gamma(\Delta_\psi+\frac{1}{2})}{ \Gamma(\Delta_{\psi}-\frac{d}{2}+1)}\frac{2^{2 \Delta _{\psi }+1} z_0^{2\Delta_\psi} z'^{2 \Delta _{\psi }-d+3}}{ \left(\left(\vec{x}'-\vec{x}_0\right){}^2+z_0^2+z'^2\right){}^{2 \Delta _{\psi }+1}}=\pi^{\frac{d-1}{2}}  \delta^{d-2}(\vec{x}'-\vec{x}_0) \delta(z'-z_0)
\ee
Therefore in the heavy limit,
\be
\begin{aligned}
\lim_{\Delta \to \infty} \delta \left(\psi(x_2) \psi(x_1) \right)&=- \frac{(u_2-u_1) \Delta_\psi}{4(4z_0^2)^{\Delta_\psi}}\int du' h_{uu}(u',v'=0,\vec{x}=\vec{x}_0,z=z_0)
\end{aligned}
\ee
This reproduces \eqref{psipsih}, which was derived in the main text using the geodesic approximation.

\subsection{Regge OPE for $(\hat{C}-2d)\delta (\psi(x_2) \psi(x_1))$}

We can also use the Witten diagram to derive the $(\hat{C}-2d)\delta (\psi(x_2) \psi(x_1))$ OPE for general operators. Using AdS invariance as in \eqref{caslight} to move the Casimir operator onto the graviton propagator, we have
\be
\begin{aligned}
&(\hat{C}_{12}-2d) \langle{O \delta (\psi \psi) O} \rangle\\
&=-(\hat{C}_{12}-2d) \int d^{d+1}x' \sqrt{-g'} \int d^{d+1}x \sqrt{-g}  T_{\alpha \beta}^{\psi,\rm bulk} G^{\alpha \beta}_{\mu \nu} T^{\mu \nu,O,\rm bulk}\\
&=-8 \pi G_N i \int d^{d+1}x' \sqrt{-g'}  T_{\alpha \beta}^{\psi,\rm bulk} T^{\alpha \beta,O,\rm bulk}
\end{aligned}
\ee
As this 4-point function calculation is true for any operator $O$, we can write down the general OPE as
\be
(\hat{C}_{12}-2d)\delta (\psi(x_2) \psi(x_1))=-8 \pi G_N i \int dx' \sqrt{-g'}  T_{\alpha \beta}^{\psi,\rm bulk} T^{\alpha \beta, \rm bulk}
\ee
where $T_{\alpha \beta}^{\psi, \rm bulk}$ is the bulk stress tensor vertex function, and $T^{\alpha \beta, \rm bulk}$ is the total bulk stress tensor operator.
Now we can use the expression \eqref{reggevertex} for $T_{uu}^{\psi,\rm bulk}$ in the Regge limit to obtain 
\be\small
\begin{aligned}
&(\hat{C}_{12}-2d) \delta (\psi(x_2) \psi(x_1))\\
&=  8\pi G_N(u_2-u_1)\int d^{d-2}\vec{x}' dz' du'  \frac{\Delta_{\psi} \Gamma(\Delta_\psi+1/2) z'^{2 \Delta_{\psi}-d+3}}{\pi^{\frac{d-1}{2}} \Gamma(\Delta_{\psi}-\frac{d}{2}+1) \Big((\vec{x}'-\vec{x}_0)^2+z_0^2+z'^2\Big)^{2\Delta_{\psi}+1}}T^{\rm bulk}_{uu}(u',v'=0,\vec{x}',z')
\end{aligned}
\ee
Taking the heavy limit using \eqref{heavylimit} gives 
\be
\lim_{\Delta_\psi \to \infty} (\hat{C}_{12}-2d) \delta (\psi(x_2) \psi(x_1))=4\pi G_N(u_2-u_1)  \frac{ \Delta_\psi}{(4z_0^2)^{\Delta_\psi}}\int du' T_{uu}^{\rm bulk}(u',v'=0,\vec{x}=\vec{x}_0,z=z_0)
\ee
This recovers \eqref{finalReggeC}, which was found in the main text by combining the geodesic approximation with the Einstein equation.

\small
\bibliographystyle{ourbst}
\bibliography{references.bib}

\end{document}